\documentclass[twocolumn]{aa}
\usepackage{graphicx}
\usepackage{txfonts}
\usepackage{lineno}

\usepackage[detect-all]{siunitx}

\begin{document} 

\title{The gamma-ray millisecond pulsar deathline, revisited}

\subtitle{New velocity and distance measurements}

\author{
L.~Guillemot$^{1,2,3}$ \and
D.~A.~Smith$^{4,5}$ \and
H.~Laffon$^{4,6}$ \and
G.~H.~Janssen$^{7}$ \and
I.~Cognard$^{1,2}$ \and
G.~Theureau$^{1,2}$ \and
G.~Desvignes$^{8}$ \and
E.~C.~Ferrara$^{9}$ \and
P.~S.~Ray$^{10}$ 
}
\institute{
\inst{1}~Laboratoire de Physique et Chimie de l'Environnement et de l'Espace -- Universit\'e d'Orl\'eans / CNRS, F-45071 Orl\'eans Cedex 02, France\\
\inst{2}~Station de radioastronomie de Nan\c{c}ay, Observatoire de Paris, CNRS/INSU, F-18330 Nan\c{c}ay, France\\
\inst{3}~email: lucas.guillemot@cnrs-orleans.fr\\
\inst{4}~Centre d'\'Etudes Nucl\'eaires de Bordeaux Gradignan, IN2P3/CNRS, Universit\'e Bordeaux 1, BP120, F-33175 Gradignan Cedex, France\\
\inst{5}~email: smith@cenbg.in2p3.fr\\
\inst{6}~email: laffon@cenbg.in2p3.fr\\
\inst{7}~ASTRON, the Netherlands Institute for Radio Astronomy, Postbus 2 7990 AA, Dwingeloo, The Netherlands\\
\inst{8}~Max-Planck-Institut f\"ur Radioastronomie, Auf dem H\"ugel 69, D-53121 Bonn, Germany\\
\inst{9}~NASA Goddard Space Flight Center, Greenbelt, MD 20771, USA\\
\inst{10}~Space Science Division, Naval Research Laboratory, Washington, DC 20375-5352, USA\\
}

\date{Received XXX; XXX}

\abstract
{Millisecond pulsars (MSPs) represent nearly half of the more than 160 currently known $\gamma$-ray pulsars detected by the Large Area Telescope on the \textit{Fermi} satellite, and a third of all known MSPs are seen in $\gamma$ rays. The least energetic $\gamma$-ray MSPs enable us to probe the so-called deathline for high-energy emission, i.e., the spin-down luminosity limit under which pulsars (PSRs) cease to produce detectable high-energy radiation. Characterizing the MSP luminosity distribution helps to determine their contribution to the Galactic diffuse $\gamma$-ray emission.}
{Because of the Shklovskii effect, precise proper motion and distance measurements are key ingredients for determining the spin-down luminosities of MSPs accurately. Our aim is to obtain new measurements of these parameters for $\gamma$-ray MSPs when possible, and clarify the relationship between the $\gamma$-ray luminosity of pulsars and their spin-down luminosity. Detecting low spin-down luminosity pulsars in $\gamma$ rays and characterizing their spin properties is also particularly interesting for constraining the deathline for high-energy emission.}
{We made use of the high-quality pulsar timing data recorded at the Nan\c{c}ay Radio Telescope over several years to characterize the properties of a selection of MSPs. For one of the pulsars, the dataset was complemented with Westerbork Synthesis Radio Telescope observations. The rotation ephemerides derived from this analysis were also used to search the LAT data for new $\gamma$-ray MSPs.}
{For the MSPs considered in this study, we obtained new transverse proper motion measurements or updated the existing ones, and placed new distance constraints for some of them, with four new timing parallax measurements. We discovered significant GeV $\gamma$-ray signals from four MSPs, i.e., PSRs J0740+6620, J0931$-$1902, J1455$-$3330, and J1730$-$2304. The latter is now the least energetic $\gamma$-ray pulsar found to date. Despite the improved $\dot E$ and $L_\gamma$ estimates, the relationship between these two quantities remains unclear, especially at low $\dot E$ values.}
{}


\maketitle

\section{Introduction}

More than 160 pulsars have now been detected as pulsed sources of GeV $\gamma$-ray emission by the Large Area Telescope (LAT), the main instrument on NASA's \textit{Fermi} Gamma-Ray Space Telescope\footnote{See \url{https://confluence.slac.stanford.edu/display/GLAMCOG/Public+List+of+LAT-Detected+Gamma-Ray+Pulsars} for an up-to-date list of $\gamma$-ray pulsar detections.}. Millisecond pulsars (MSPs), neutron stars with very short rotational periods ($P \lesssim 30$ ms) that have been spun up by accretion from a binary companion \citep{Alpar1982,Bhattacharya1991,Tauris2006} currently represent over 40\% of this total. In addition to those already detected as pulsed sources of $\gamma$ rays, two dozen more MSPs were discovered in targeted radio searches at the locations of unassociated \textit{Fermi} LAT sources with pulsar-like properties, and will likely be confirmed as the sources of the high-energy emission, once accurate timing models that are valid over months to years are available for them. A recent systematic study of the temporal and spectral emission properties of pulsars in $\gamma$ rays is presented in the Second \textit{Fermi} LAT Pulsar Catalog \citep[hereafter 2PC;][]{Fermi2PC}, and \citet{Johnson2014} carried out a systematic modeling analysis of MSP radio and $\gamma$-ray light curves in the context of a selection of geometric emission models. 

As the \textit{Fermi} mission continues, the accumulation of $\gamma$-ray data by the LAT enables us to detect pulsations from fainter and fainter pulsars. This is especially true with the advent of the Pass 8 data, which is based on much improved LAT event reconstruction algorithms \citep{Atwood2013}, increasing the sensitivity of the LAT to $\gamma$-ray pulsations. As is shown in, for example, 2PC, the $\gamma$-ray luminosity of pulsars $L_\gamma \propto h d^2$, where $h$ is the $\gamma$-ray energy flux and $d$ is the distance, scales with the so-called spin-down luminosity $\dot E = 4 \pi^2 I \dot P / P^3$, where $I$ denotes the neutron star moment of inertia (generally, and here, assumed to be $10^{45}$ g cm$^2$, corresponding to canonical mass and radius values of 1.4 M$_\odot$ and 10 km respectively), $P$ is the rotational period and $\dot P$ is the spin-down rate. Increased sensitivity to $\gamma$-ray pulsations thus enables us to probe greater distances as well as smaller $\dot E$ values, i.e., less and less energetic pulsars. Observing low $\dot E$ objects in $\gamma$ rays allows us to constrain the deathline for $\gamma$-ray emission, which we define as the spin-down luminosity limit under which pulsars can no longer produce detectable high-energy radiation. Prior to this work, the empirical deathline was $\dot E_\mathrm{death} \sim 3 \times 10^{33}$ erg s$^{-1}$ (see 2PC). This deathline is a key unknown for models of high-energy emission from pulsars.

The present study focuses on MSPs which, among several other differences, are more widely distributed in Galactic latitudes than normal, non-recycled pulsars. Because of the bright $\gamma$-ray diffuse background at low Galactic latitudes, faint normal pulsars are more difficult to observe than faint MSPs. The sample of low $\dot E$ objects is thus likely more complete among $\gamma$-ray MSPs than normal $\gamma$-ray pulsars. Additionally, as the sample of $\gamma$-ray pulsars grows, the MSP pulse profile and luminosity distributions appear increasingly to differ from those of the normal pulsars, warranting that the two populations be studied separately.

One difficulty in probing the high-energy emission deathline for MSPs is that low $\dot E$ MSPs tend to have very small spin-down rates, typically of the order of $\dot P \sim 10^{-21}$ s/s (we omit these units when quoting spin-down rates in the remainder of the text). Such low spin-down rates can be strongly affected by kinematic effects causing the apparent (measured) values, $\dot P_\mathrm{obs}$, to differ substantially from the intrinsic spin-down rates, $\dot P_\mathrm{int}$. In turn, this can lead to spin-down luminosity values that are bad estimates of the true energy budget that can be converted into $\gamma$-ray emission. The intrinsic spin-down rate can be expressed as

\begin{eqnarray}
\dot P_\mathrm{int} = \dot P_\mathrm{obs} - \dot P_\mathrm{Gal} - \dot P_\mathrm{Shk}.
\end{eqnarray}

In the above expression, $\dot P_\mathrm{Gal}$ is the difference between the line of sight acceleration of the pulsar and the Solar System in the gravitational potential of the Galaxy (see for example 2PC for further details), and $\dot P_\mathrm{Shk}$ denotes the kinematic Shklovskii correction caused by the pulsar's transverse proper motion \citep{Shklovskii1970}, calculated as

\begin{eqnarray}
\dot P_\mathrm{Shk} \simeq 2.43 \times 10^{-21} \left( \frac{\mu_\perp}{\mathrm{mas\ yr}^{-1}} \right)^2 \left( \frac{d}{1\ \mathrm{kpc}} \right) \left( \frac{P}{\mathrm{s}} \right), 
\end{eqnarray}

\noindent
where $\mu_\perp$ denotes the transverse proper motion of the pulsar. We note that for pulsars undergoing line of sight acceleration in an external gravitational field, as is for instance commonly observed for pulsars in globular clusters, the spin-down rate is shifted by $a_l P / c$, where $a_l$ is the acceleration along the line of sight. From the above expressions it is clear that accurate spin-down rate, proper motion, and distance estimates are important for properly characterizing our $\gamma$-ray MSPs, and especially those near the $\gamma$-ray emission deathline that are most sensitive to inaccuracies in $\dot P$ corrections. We note that varying moment of inertia ($I$) values could also play a role in the $L_\gamma$ versus $\dot E$ relationship. In future analyses it may be worthwhile to account for the influence of the moment of inertia on this relationship. 

The present article reports on the analysis of radio pulsar timing data taken at the Nan\c{c}ay Radio Telescope (NRT) for a selection of $\gamma$-ray MSPs with no or incomplete proper motion information. We obtained solid proper motion parameters for the selected pulsars, and new distance estimates from timing parallaxes for a few of them. With the new proper motion and distance data we then derived new spin-down luminosity $\dot E$ estimates and reassessed the question of the high-energy emission deathline. We also searched the data recorded by the LAT for pulsations using pulsar timing information from the NRT, and discovered pulsed GeV emission from four MSPs: PSRs J0740+6620, J0931$-$1902, J1455$-$3330, and J1730$-$2304. In Section~\ref{sec:radiotiming}, we present the list of pulsars selected for the radio timing analysis with NRT data, and present the results of this analysis. The analysis of LAT $\gamma$-ray data for PSRs J0740+6620, J0931$-$1902, J1455$-$3330, and J1730$-$2304 is reported in section~\ref{sec:gamma}. The latter sections are followed by a discussion of the detectability of energetic MSPs in $\gamma$ rays, the deathline for $\gamma$-ray emission from MSPs and the relationship between $L_\gamma$ and $\dot E$. Section~\ref{sec:summary} summarizes our results.

\section{Radio timing analysis}
\label{sec:radiotiming}

\subsection{Pulsar selection}
\label{sec:pulsarsec} 

The list of MSPs considered in the study was built by selecting those with significant $\gamma$-ray pulsations detected with the \textit{Fermi} LAT, and that are also regularly observed at the NRT. We further selected MSPs with no or incomplete transverse proper motion ($\mu_\perp$) information in version 1.51 of the Australian Telescope National Facility (ATNF) pulsar database\footnote{\url{http://www.atnf.csiro.au/people/pulsar/psrcat/}} \citep{ATNF}, i.e., with either the `PMRA' or the `PMDEC' parameter missing, or with the `PMTOT' parameter available but both `PMRA' and `PMDEC' missing. Pulsars with complete proper motion information available were therefore rejected.

A few pulsars of interest were added manually to the list obtained after the selection described above. PSRs J0610$-$2100, J1024$-$0719 and J1231$-$1411 are cases of objects with large Shklovskii corrections causing very low or negative $\dot E$ values, indicating possible issues with their proper motion and distance estimates \citep[see discussions in 2PC;][]{Espinoza2013,GuillemotTauris2014}. These three pulsars were added to our list of targets, with the hope of shedding light on the causes of the likely overestimated Shklovskii corrections. Although the MSP discovered at Nan\c{c}ay in a \textit{Fermi} LAT unassociated source PSR~J2043+1711 has a published proper motion measurement \citep{gfc+12}, the much increased radio timing data span since the latter publication was likely to provide improved constraints on $\mu_\perp$. Finally, the four new $\gamma$-ray MSP detections PSRs J0740+6620, J0931$-$1902, J1455$-$3330, and J1730$-$2304 were included in the study, to obtain pulsar timing parameters enabling precise phase-folding of the LAT data (see Section~\ref{sec:gamma}). 

Table~\ref{tab:MSPs} lists the 19 MSPs retained for our study and some of their properties. Nine distances come from timing parallax measurements, of which seven come from this work, described in the next section. The remaining distances come from the electron column density along the line of sight to the pulsar (the dispersion measure, DM) and the NE2001 Galactic free electron model \citep{NE2001}. The DM distance uncertainties were estimated by varying the DM by $\pm$ 20\% as in 2PC. For each line of sight, we examined the NE2001 electron density versus distance. The density is highly structured for most of these pulsars, with unphysical steps, dips, and spikes, especially for the first few kpc: see Figure 4 of \citet{Hou2014} for examples. The DM value is small ($<$ 25 pc cm$^{-3}$) for half of these MSPs, making the estimated distance particularly sensitive to the local density model, with uncertainties perhaps larger than those tabulated.

\begin{table*}
\caption{
Main properties of the MSPs in this study. The columns give the pulsar names, Galactic coordinates, rotational periods, apparent spin-down rates, distances, corrected spin-down rates and spin-down powers, and the $>$ 100 MeV $\gamma$-ray energy fluxes, luminosities and efficiencies. Numbers in parentheses are 1$\sigma$ uncertainties in the last digit(s) quoted. Distance values marked with a \dag\ symbol are derived from the pulsar DM, with uncertainties as described in the text. The distance for PSR~J1741+1351 (marked with a $\star$) comes from \citet{Espinoza2013}, while that for PSR~J1823$-$3021A (marked with a \ddag\ symbol) is from \citet{Kuulkers2003}. Other distance values are from timing parallax measurements obtained in this study (see Section~\ref{sec:radioresults}). The $>$ 100 MeV $\gamma$-ray energy fluxes are taken from the 3FGL catalog \citep{Fermi3FGL}, except for PSR~J1811$-$2405 \citep{nbb+14}, and the values marked with a $\Vert$ for which the fluxes were determined here (see Section~\ref{sec:gamma}).
}
\label{tab:MSPs}
\centering
\begin{scriptsize}
\begin{tabular}{c S[table-format=3.2] S[table-format=3.2] S[table-format=1.4] S[table-format=3.8] S[table-format=1.5] S[table-format=3.8] S[table-format=3.6] S[table-format=2.5] S[table-format=3.4] S[table-format=3.4]}
\hline
\hline
Pulsar        & \multicolumn{1}{c}{$l$}        & \multicolumn{1}{c}{$b$}        & \multicolumn{1}{c}{$P$}   & \multicolumn{1}{c}{$\dot P_\mathrm{obs}$} & \multicolumn{1}{c}{$d$}            & \multicolumn{1}{c}{$\dot P_\mathrm{int}$} & \multicolumn{1}{c}{$\dot E_\mathrm{int}$}    & \multicolumn{1}{c}{$h$}                                 & \multicolumn{1}{c}{$L_\gamma$}               & \multicolumn{1}{c}{$\eta$}   \\
              & \multicolumn{1}{c}{($^\circ$)} & \multicolumn{1}{c}{($^\circ$)} & \multicolumn{1}{c}{(ms)}  & \multicolumn{1}{c}{($10^{-20}$)}          & \multicolumn{1}{c}{(kpc)}          & \multicolumn{1}{c}{($10^{-20}$)}          & \multicolumn{1}{c}{($10^{33}$ erg s$^{-1}$)} & \multicolumn{1}{c}{($10^{-11}$ erg cm$^{-2}$ s$^{-1}$)} & \multicolumn{1}{c}{($10^{33}$ erg s$^{-1}$)} & \multicolumn{1}{c}{\%}       \\
\hline
J0034$-$0534  & 111.49 & -68.07   & 1.877 & 0.49762(9)            & 0.54(11)\dag   & 0.489(12)             & 29.2(7)                  & 1.80(10)                            & 0.6(3)                   & 2(1)     \\
J0340+4130    & 153.78 & -11.02   & 3.299 & 0.70486(15)           & 1.73(30)\dag   & 0.661(5)              & 7.27(6)                  & 2.22(13)                            & 8(3)                     & 110(40)  \\
J0610$-$2100  & 227.75 & -18.18   & 3.861 & 1.23317(4)            & 3.5(10)\dag    & 0.01(34)              & 0.1(23)                  & 1.15(11)                            & 17(10)                   & \multicolumn{1}{c}{--}       \\
J0614$-$3329  & 240.50 & -21.83   & 3.149 & 1.74169(8)            & 1.9(4)\dag     & 1.7616(15)            & 22.279(19)               & 11.10(24)                           & 48(22)                   & 220(100) \\
J0740+6620    & 149.73 & 29.60      & 2.886 & 1.23(6)               & 0.68(10)\dag   & 0.73(16)              & 12.0(26)                 & 0.36(23)$\Vert$                     & 0.2(1)                   & 2(1) \\
J0751+1807    & 202.73 & 21.09      & 3.479 & 0.778802(9)           & 1.51(35)       & 0.53(6)               & 5.0(5)                   & 1.30(10)                            & 4(2)                     & 70(40)   \\
J0931$-$1902 & 251.00 & 23.05     & 4.638 & 0.3612(17)               & 1.88(51)\dag & 0.37(3)               & 1.5(1)                      & 0.67(22)$\Vert$                     & 2.8(18)                & 200(130) \\
J1024$-$0719  & 251.70 & 40.52      & 5.162 & 1.855242(27)          & 1.13(18)       & -3.1(8)             & -8.9(23)               & 0.36(5)                             & 0.5(2)                   & \multicolumn{1}{c}{--}       \\
J1231$-$1411  & 295.53 & 48.39      & 3.684 & 2.26235(11)           & 0.44(5)\dag    & 0.77(17)              & 6.1(14)                  & 10.30(21)                           & 2.4(5)                   & 39(18)   \\
J1455$-$3330  & 330.72 & 22.56      & 7.987 & 2.43072(5)            & 1.01(22)       & 2.330(28)             & 1.806(22)                & 0.42(15)$\Vert$                     & 0.5(3)                   & 28(16)   \\
J1614$-$2230  & 352.64 & 20.19      & 3.151 & 0.962464(11)          & 0.77(5)        & 0.34(5)               & 4.3(6)                   & 2.33(15)                            & 1.6(2)                   & 38(10)   \\
J1730$-$2304  & 3.14 & 6.02       & 8.123 & 2.018340(28)          & 0.84(19)       & 1.15(29)              & 0.84(22)                 & 1.0(5)$\Vert$                     & 1.1(10)                   & 130(130) \\
J1741+1351    & 37.89 & 21.64      & 3.747 & 3.02163(6)            & 1.08(5)$\star$ & 2.909(7)              & 21.82(5)                 & 0.57(11)                            & 0.8(2)                   & 4(1) \\
J1811$-$2405  & 7.07 & -2.56    & 2.661 & 1.33708(7)            & 1.8(5)\dag     & 1.20(11)              & 25.1(23)                 & 1.4(8)                              & 5(4)                     & 20(17)   \\
J1823$-$3021A & 2.79 & -7.91    & 5.440 & 337.61901(32)         & 7.6(4)\ddag    & 336.42(28)            & 825.0(7)                 & 1.59(17)                            & 110(20)                  & 13(2)    \\
J2017+0603    & 48.62 & -16.03   & 2.896 & 0.79936(7)            & 0.9(4)         & 0.8122(15)            & 13.199(24)               & 3.49(17)                            & 3(3)                     & 23(20)   \\
J2043+1711    & 61.92 & -15.31   & 2.380 & 0.52480(21)           & 1.76(32)\dag   & 0.393(32)             & 11.5(9)                  & 3.02(14)                            & 11(4)                    & 100(40)  \\
J2214+3000    & 86.86 & -21.67   & 3.119 & 1.47269(9)            & 0.60(31)       & 1.30(10)              & 16.9(13)                 & 3.30(12)                            & 1.5(15)                 & 9(9)     \\
J2302+4442    & 103.40 & -14.00   & 5.192 & 1.38689(17)           & 1.19(23)\dag   & 1.386(10)             & 3.908(29)                & 3.81(14)                            & 6.5(25)                 & 170(70)  \\
\hline
\end{tabular}
\end{scriptsize}
\end{table*}

\subsection{Methodology}

The radio pulsar timing analysis was carried out by analyzing data recorded with the Berkeley-Orl\'eans-Nan\c{c}ay (BON) and the NUPPI (a version of the Green Bank Ultimate Pulsar Processing Instrument\footnote{https://safe.nrao.edu/wiki/bin/view/CICADA/NGNPP} designed for the NRT) instruments in operation at the NRT, a meridian telescope equivalent to a 94 m dish located near Orl\'{e}ans, France. The BON backend was the main pulsar timing instrument at Nan\c{c}ay after it started operating in late 2004. BON observes primarily at 1.4 GHz, with a frequency bandwidth of 64 MHz before July 2008 and then 128 MHz. Data recorded with this instrument are coherently dedispersed within 4 MHz channels, ensuring good timing resolution. In August 2011, BON was replaced as the principal instrument for pulsar observations by NUPPI, a new backend giving access to a much increased frequency bandwidth of 512 MHz, also coherently dedispersed. NUPPI pulsar observations are primarily carried out at 1.4 GHz. The BON backend is still active and is used in parallel to NUPPI, with a central frequency for BON observations moved up to 1.6 GHz while NUPPI observations remain centered at 1.4 GHz. Because the frequency band recorded by the BON backend after August 2011 overlaps with the one covered by NUPPI, we excluded from our datasets BON observations made simultaneously with NUPPI observations, to avoid duplication of astrophysical information. As a result, and since no 1.4 GHz BON data were available for PSR J0740+6620, only NUPPI data were analyzed for this pulsar.

The subsequent data reduction was done using the PSRCHIVE analysis software library \citep{PSRCHIVE}. We cleaned the data for the selected MSPs of radio frequency interference (RFI) using a median smoothed automatic channel zapping algorithm as implemented in PSRCHIVE, and polarization-calibrated the data with the \textsc{SingleAxis} method of the software package. For each pulsar and observation system (BON 1.4 GHz and NUPPI 1.4 GHz), we combined up to 10 observations with the highest signal-to-noise ratios (S/N) to produce high-S/N integrated profiles. These profiles were smoothed to produce template profiles, and times of arrival (TOAs) were then extracted by cross-correlating the template profiles with the observations. The TOA extraction was carried out using the ``Fourier phase gradient algorithm'' that uses the property that two functions shifted in the time domain have Fourier transforms that differ by a linear phase gradient \citep[a detailed description of the method can be found in][]{Taylor1992}, except for the pulsars with generally low S/N profiles PSRs J0034$-$0534 and J2043+1711, for which the ``Fourier domain with Markov chain Monte Carlo algorithm'' gave more realistic TOA uncertainties. The TOA data were analyzed with the \textsc{Tempo2} pulsar timing software \citep{TEMPO2}. A detailed description of pulsar timing equations and techniques can be found in, e.g., \citet{handbook}. We used the DE421 Solar system ephemeris \citep{Folkner2009} for the conversion of the TOAs to Barycentric Coordinate Time (TCB). 

In a first iteration of the timing analysis, the BON observations were split into two frequency channels of 32 MHz before July 2008 and of 64 MHz after that date, and the NUPPI observations were split into four frequency channels of 128 MHz each. The multi-frequency TOA datasets created with this procedure allowed us to track potential time variations of the integrated column density of free electrons along the line of sight, the DM. Using \textsc{Tempo2} and the TOA datasets, we obtained new timing solutions for each pulsar by fitting for their astrometric parameters (equatorial coordinates and proper motion components), rotational parameters (rotational frequency and first time derivative), and binary parameters when applicable. We also fitted for the DM value and for its first time derivative (`DM1' parameter in \textsc{Tempo2}), excluding the latter parameter from the model if not significant. We fitted a systematic time offset (`JUMP' parameter) between the BON and the NUPPI TOA datasets, in order to accommodate differences in the observing systems and in the template profiles. The timing solutions obtained after this first iteration were then used to phase-fold the observation files, and new integrated profiles and template profiles were created with the updated observation files. 

In the second iteration of the analysis, we concatenated the frequency information from the 1.4 GHz BON and NUPPI data to produce one TOA per observation, representing the entire frequency bandwidth available. The DM parameters were frozen at the values obtained from the previous step, and the timing analysis was repeated. To clean the TOA residual data of any remaining outliers degrading the timing analysis, we rejected residuals $r_i$ verifying $\left| r_i - \mathrm{med}(r) \right| > K \sigma$, where $\mathrm{med}(r)$ denotes the median value of the residuals, $\sigma$ is the median absolute deviation \citep[MAD; see for example][]{Huber1981}, and setting $K$ to 3 which approximates a cut at two standard deviations for a Gaussian distribution. 

In the case of PSR~J0931$-$1902, the Nan\c{c}ay timing dataset was complemented by adding Westerbork Synthesis Radio Telescope (WSRT) TOAs generated from the PuMa-II backend \citep{Karuppusamy2008}. Observations for this pulsar were done on a monthly basis at two frequencies using the 1380 MHz receiver with a bandwidth of 160 MHz and  the 350 MHz receiver with a bandwidth of 80 MHz. All data were coherently dedispersed using \textsc{dspsr} and folded using the PSRCHIVE software in a similar way as the Nan\c{c}ay data. To generate TOAs, synthetic templates were generated based on high S/N additions of all available observations at each frequency. The extended timing baseline and the quality of the WSRT TOAs improved the measurement of the astrometric parameters. 

Table~\ref{table:radiodata} summarizes the properties of the TOAs selected at this stage of the analysis. The main results from the timing analysis with \textsc{Tempo2} are summarized in Section~\ref{sec:radioresults}.

\begin{table*}
\caption{
Properties of the radio timing data analyzed in this study. For each pulsar and observation
backend, we list the number of selected TOAs, the MJD range, the time span of the TOA datasets,
and the median uncertainty of the individual TOAs.
}
\label{table:radiodata}
\centering
\begin{tabular}{c c | c c c c c}
\hline
\hline
Pulsar Name & Residual RMS ($\mu$s) & Backend & N$_\mathrm{TOA}$ & MJD Range & Time span (yrs) & $\sigma_\mathrm{med}$ ($\mu$s) \\
\hline

J0034$-$0534 & 7.99 & BON & 66 & 53761.7 --- 55808.1 & 5.6 & 21.02 \\
             &   & NUPPI & 30 & 55935.7 --- 57032.7 & 3.0 & 16.13 \\
J0340+4130 & 3.31 & BON & 170 & 55309.6 --- 55862.0 & 1.5 & 2.73 \\
           &   & NUPPI & 143 & 55823.2 --- 56805.5 & 2.7 & 2.10 \\
J0610$-$2100 & 2.19 & BON & 89 & 54270.5 --- 55805.3 & 4.2 & 2.60 \\
             &   & NUPPI & 49 & 55854.2 --- 57047.9 & 3.3 & 1.40 \\
J0614$-$3329 & 1.53 & BON & 60 & 55160.1 --- 55857.2 & 1.9 & 2.20 \\
             &   & NUPPI & 54 & 55838.2 --- 57038.9 & 3.3 & 1.69 \\
J0740+6620 & 0.46 & NUPPI & 43 & 56675.0 --- 57037.0 & 1.0 & 1.05 \\
J0751+1807 & 1.17 & BON & 198 & 53373.0 --- 55880.2 & 6.9 & 2.05 \\
           &   & NUPPI & 153 & 55825.3 --- 57047.0 & 3.3 & 0.83 \\
J0931$-$1902 & 4.76 & NUPPI & 27 & 56399.8 --- 57044.0 & 1.8 & 4.98 \\
             &   & PuMa-II (1.4 GHz) & 23 & 56113.7 --- 57061.9 & 2.6 & 10.29 \\
             &   & PuMa-II (0.35 GHz) & 32 & 56113.7 --- 57118.8 & 2.8 & 10.51 \\
J1024$-$0719 & 0.96 & BON & 184 & 53714.2 --- 55807.5 & 5.7 & 1.65 \\
             &   & NUPPI & 118 & 55819.5 --- 57047.1 & 3.4 & 1.12 \\
J1231$-$1411 & 4.42 & BON & 174 & 55168.3 --- 55878.4 & 1.9 & 3.62 \\
             &   & NUPPI & 207 & 55877.4 --- 57046.2 & 3.2 & 3.40 \\
J1455$-$3330 & 1.75 & BON & 248 & 54238.0 --- 55881.5 & 4.5 & 4.63 \\
             &   & NUPPI & 260 & 55819.6 --- 57049.3 & 3.4 & 2.94 \\
J1614$-$2230 & 0.47 & BON & 80 & 54896.2 --- 55881.5 & 2.7 & 0.65 \\
             &   & NUPPI & 142 & 55853.6 --- 57032.4 & 3.2 & 0.45 \\
J1730$-$2304 & 1.22 & BON & 113 & 53385.4 --- 55852.7 & 6.8 & 2.15 \\
             &   & NUPPI & 47 & 55923.5 --- 57047.4 & 3.1 & 0.65 \\
J1741+1351 & 1.21 & BON & 38 & 54085.5 --- 55903.5 & 5.0 & 2.24 \\
           &   & NUPPI & 18 & 55812.8 --- 57051.4 & 3.4 & 1.58 \\
J1811$-$2405 & 0.48 & BON & 4 & 55597.4 --- 55735.0 & 0.4 & 0.46 \\
             &   & NUPPI & 43 & 55871.6 --- 57048.4 & 3.2 & 0.37 \\
J1823$-$3021A & 3.77 & BON & 28 & 53784.3 --- 55889.6 & 5.8 & 4.47 \\
              &   & NUPPI & 22 & 55980.3 --- 57038.4 & 2.9 & 2.44 \\
J2017+0603 & 1.22 & BON & 50 & 55246.4 --- 55871.7 & 1.7 & 2.34 \\
           &   & NUPPI & 57 & 55879.7 --- 57048.5 & 3.2 & 1.60 \\
J2043+1711 & 1.19 & BON & 23 & 55425.0 --- 55841.8 & 1.1 & 2.56 \\
           &   & NUPPI & 22 & 55877.7 --- 57029.6 & 3.2 & 2.62 \\
J2214+3000 & 2.54 & BON & 98 & 55136.8 --- 55856.8 & 2.0 & 2.82 \\
           &   & NUPPI & 78 & 55819.9 --- 56954.8 & 3.1 & 1.94 \\
J2302+4442 & 2.57 & BON & 94 & 55150.8 --- 55869.8 & 2.0 & 3.99 \\
           &   & NUPPI & 93 & 55852.9 --- 57047.6 & 3.3 & 2.19 \\

\hline
\end{tabular}
\end{table*}

\subsection{Results}
\label{sec:radioresults} 

The post-fit timing residual root-mean-square (rms, given by the `TRES' parameter in \textsc{Tempo2}) values for each MSP considered in this study are given in Table~\ref{table:radiodata}, along with a summary of the main properties of the NRT TOAs. Reduced $\chi^2$ values for the MSPs in the table range from 1.0 to 2.5, indicating that the timing solutions describe the TOAs adequately. For all pulsars, the timing precision and time interval considered enabled us to measure the proper motions. In a few cases we could also detect an annual parallax in the TOA residuals. For a pulsar at a distance $d$ and with an ecliptic latitude of $\beta$, the parallax effect introduces a sinusoidal variation in the TOA residuals with an amplitude of $l^2 \cos^2(\beta) / \left(2 c d\right)$, where $l$ is the Earth-Sun distance, and $c$ is the speed of light. The effect is subtle: at $d = 1$ kpc and for $\beta = 0$ the amplitude is only 1.2 $\mu$s; so it was only measurable for a subset of the MSPs with low residual rms values.

The proper motion and parallax measurements are listed in Table~\ref{tab:pm_px} along with the associated 1$\sigma$ uncertainties from \textsc{Tempo2}. Also given in the table are the previously reported values when available, for comparison. For the MSPs with no parallax measurement we determined 2$\sigma$ upper limits; these limits are also reported in the table. Our PM and PX values are consistent with the latest results from the EPTA collaboration \citep{Desvignes2016} who combined data recorded with the BON backend, analyzed with different methods, and from other radio telescopes. Our measurements are also compatible with those published by the NANOGrav and PPTA collaborations \citep{Matthews2015,Reardon2016}.

We explored how the Lutz-Kelker effect \citep{LutzKelker1973} changes the inferred pulsar distances, using the code provided by \citet{Verbiest2012}, for the six pulsars for which we measured a timing parallax. For four of the pulsars, the ATNF database lists values of the flux density at 1400 MHz, which we provided to the code. The distance is decreased for all six pulsars. For four of the pulsars the corrected distance is within 15\% of PX$^{-1}$, with no consequences for our conclusions. For PSRs J2017+0603 and J2214+3000 the distances decrease to 0.4 and 0.2 kpc, respectively, i.e., about 40\% of the uncorrected values. Even this rather significant change does not qualitatively change our conclusions (e.g., for the $\gamma$-ray luminosity and efficiency values; see the paragraphs for these two pulsars below). We neglect the Lutz-Kelker effect in the rest of this paper.

The new proper motion parameters as well as the distances derived from the parallax measurements were used to calculate the Shklovskii corrections to $\dot P$. The distances of pulsars with no detection of the parallax signature were determined using the NE2001 model of Galactic free electron density \citep{NE2001}. The spin-down rate values corrected for the Shklovskii effect and for the acceleration in the Galactic potential are given in Table~\ref{tab:MSPs}. In the following we present the salient results stemming from the timing analysis, and the new spin-down rate estimates. The cases of PSRs~J0610$-$2100 and J1024$-$0719 are discussed in separate sections, \ref{sec:J0610-2100} and \ref{sec:J1024-0719}. \\

\begin{table*}
\caption{
Proper motion and parallax measurements for the pulsars considered in this study. Quoted uncertainties on the measured parameters PMRA, PMDEC, and PX, are the 1$\sigma$ statistical error bars from \textsc{Tempo2}. In the cases where the parallax was not measurable, we quote 2$\sigma$ upper limits based on the radio timing data. For pulsars with previously reported proper motion or parallax values, we report these results and give the associated references. References: (1) -- \citet{hllk05}, (2) -- \citet{bjd+06}, (3) -- \citet{nss+05}, (4) -- \citet{vbc+09}, (5) -- \citet{hbo06}, (6) -- \citet{rrc+11}, (7) -- \citet{tsb+99}, (8) -- \citet{bk11}, (9) -- \citet{dpr+10}, (10) -- \citet{Espinoza2013}, (11) -- \citet{nbb+14}, (12) -- \citet{gfc+12}
}
\label{tab:pm_px}
\centering
\begin{scriptsize}
\begin{tabular}{c | S[table-format=3.6] S[table-format=4.5] | S[table-format=4.6] S[table-format=3.4] | S[table-format=4.5] S[table-format=3.4] | S[table-format=2.5,table-space-text-pre=<] S[table-format=1.4] | S[table-format=2.5,table-space-text-pre=>] S[table-format=1.4] | c}
\hline
\hline
Pulsar & \multicolumn{2}{c|}{PMRA (mas yr$^{-1}$)} & \multicolumn{2}{c|}{PMDEC (mas yr$^{-1}$)} & \multicolumn{2}{c|}{PMTOT (mas yr$^{-1}$)} & \multicolumn{2}{c|}{PX (mas)} & \multicolumn{2}{c|}{Derived PX distance (kpc)} & References \\
 & {This work} & {Prev.} & {This work} & {Prev.} & {This work} & {Prev.} & {This work} & {Prev.} & {This work} & {Prev.} & \\
\hline
J0034$-$0534  & 7.9(8)      & {--}         & -9.9(17)  & {--}         & 12.6(14) & 31(9)    & {<} 7.4     & {--}        & {>} 0.14    & {--}         & --, --, 1, --  \\
J0340+4130    & -0.59(16)   & {--}         & -3.81(34)  & {--}         & 3.85(33)  & {--}         & {<} 1.3     & {--}        & {>} 0.77    & {--}         & --, --, --, -- \\
J0610$-$2100  & 9.21(6)     & 7(3)     & 16.73(8)   & 11(3)    & 19.10(8)  & 13(3)    & {<} 1.3     & {--}        & {>} 0.77    & {--}         & 2, 2, 2, --    \\
J0614$-$3329  & 0.58(9)     & {--}         & -1.92(12)  & {--}         & 2.00(11)  & {--}         & {<} 2.2     & {--}        & {>} 0.45    & {--}         & --, --, --, -- \\
J0740+6620    & -6(11)  & {--}         & -32(4) & {--}         & 32.6(41) & {--}         & {<} 11.7    & {--}        & {>} 0.09    & {--}         & --, --, --, -- \\
J0751+1807    & -2.71(7)    & {--}         & -13.2(4)   & {--}         & 13.51(35) & 6.0(20)  & 0.66(15) & 1.6(8)  & 1.51(35) & 0.62(31) & --, --, 3, 3   \\
J0931$-$1902  & -1.1(8)     & {--}         & -4.4(12)  & {--}         & 4.6(12)  & {--}         & {<} 5.0     & {--}        & {>} 0.20    & {--}         & --, --, --, -- \\
J1024$-$0719  & -35.247(23) & -35.3(2) & -48.14(5)  & -48.2(3) & 59.67(4)  & 59.7(3)  & 0.89(14) & 1.9(8)  & 1.13(18) & 0.53(22) & 4, 4, 4, 5     \\
J1231$-$1411  & -62.03(26)  & -100(20) & 6.2(5)     & -30(40)  & 62.34(26) & 104(22)  & {<} 1.8     & {--}        & {>} 0.56    & {--}         & 6, 6, 6, --    \\
J1455$-$3330  & 7.88(5)     & 5(6)     & -1.90(12)  & 24(12)   & 8.11(5)   & 25(12)   & 0.99(22) & {--}        & 1.01(22) & {--}         & 7, 7, 7, --    \\
J1614$-$2230  & 3.87(12)    & {--}         & -32.3(7)   & {--}         & 32.5(6)   & 32(3)    & 1.30(9)  & 0.5(6)  & 0.77(5)  & 2.0(24) & --, --, 8, 9   \\
J1730$-$2304  & 20.7(4)     & 20.27(6) & 8.3(83)   & {--}         & 22.3(31) & {--}         & 1.19(27) & {--}        & 0.84(19) & {--}         & 4, --, --, --  \\
J1741+1351    & -8.93(8)    & {--}         & -7.43(17)  & {--}         & 11.62(13) & 11.71(1) & {<} 1.2     & 0.93(4) & {>} 0.83    & 1.08(5)  & --, --, 10, 10 \\
J1811$-$2405  & 0.65(14)    & 0.53(13) & -9.1(52)  & {--}         & 9.2(51)  & {--}         & {<} 0.4     & {--}        & {>} 2.50    & {--}         & 11, --, --, -- \\
J1823$-$3021A & 0.31(24)    & {--}         & -8.2(17)  & {--}         & 8.2(17)  & {--}         & {<} 7.0     & {--}        & {>} 0.14    & {--}         & --, --, --, -- \\
J2017+0603    & 2.35(8)     & {--}         & 0.17(16)   & {--}         & 2.35(8)   & {--}         & 1.2(5)   & {--}        & 0.9(4)   & {--}         & --, --, --, -- \\
J2043+1711    & -6.12(27)   & -7(2)    & -11.2(5)   & -11(2)   & 12.8(4)   & 13(2)    & {<} 4.4     & {--}        & {>} 0.23    & {--}         & 12, 12, 12, -- \\
J2214+3000    & 20.90(11)   & {--}         & -1.55(15)  & {--}         & 20.96(11) & {--}         & 1.7(9)   & {--}        & 0.60(31) & {--}         & --, --, --, -- \\
J2302+4442    & -0.05(13)   & {--}         & -5.85(12)  & {--}         & 5.85(12)  & {--}         & {<} 2.5     & {--}        & {>} 0.40    & {--}         & --, --, --, -- \\

\hline
\end{tabular}
\end{scriptsize}
\end{table*}

\begin{itemize}

\item[$\bullet$] \textit{PSR~J0034$-$0534:} nearly nine years of NRT data yield a total transverse proper motion of $\mu_\perp = (12.6 \pm 1.4)$ mas yr$^{-1}$, significantly smaller than the value of $(31 \pm 9)$ mas yr$^{-1}$ determined by \citet{hllk05} and thus reducing the Shklovskii correction appreciably: our estimate of $\dot P_\mathrm{int}$ is about 70\% larger than the value of $\sim 2.9 \times 10^{-21}$ reported in 2PC, assuming the same NE2001 distance of $(0.54 \pm 0.11)$ kpc. The pulsar's efficiency of conversion of spin-down power into $\gamma$ radiation, $\eta = L_\gamma / \dot E_\mathrm{int}$, decreases slightly from about 3\% to 2\%. \\

\item[$\bullet$] \textit{PSR~J0340+4130:} the modest proper motion determined for this pulsar of $(3.85 \pm 0.33)$ mas yr$^{-1}$ introduces a small Shklovskii correction to the observed spin-down rate value at the NE2001 distance of $(1.73 \pm 0.30)$ kpc. A $\gamma$-ray luminosity of about $7.3 \times 10^{33}$ erg s$^{-1}$ was determined in 2PC for this pulsar. The slightly reduced $\dot E_\mathrm{int}$ value compared to that quoted in 2PC makes the $\gamma$-ray efficiency $\eta$ to be $\sim$110\%. PSR~J0340+4130 is likely closer than the NE2001 distance of 1.73 kpc, which would reduce $L_\gamma$ significantly and also diminish the Shklovskii correction. No significant timing parallax is detected for this pulsar with the present NRT dataset. \\

\item[$\bullet$] \textit{PSR~J0614$-$3329:} this pulsar is listed in 2PC as having an implausible $\gamma$-ray efficiency value of about 215\%, at the NE2001 distance of $(1.9 \pm 0.4)$ kpc. Interestingly, we measure a very modest transverse proper motion of $\mu_\perp = (2.00 \pm 0.11)$ mas yr$^{-1}$ for this MSP, leading to a negligible Shklovskii correction. No parallax measurement was possible with the Nan\c{c}ay timing data. Yet, the pulsar may lie at as little as a quarter of the NE2001 distance, given its high $\gamma$-ray efficiency. \\

\item[$\bullet$] \textit{PSR~J0740+6620:} this pulsar, named PSR~J0741+66 in \citet{Stovall2014} and PSR~J0742+66 in the ATNF pulsar database, is found to be at right ascension $\alpha_\mathrm{J2000} = 07^h$$40^m$$45.798(5)^s$ and declination $\delta_\mathrm{J2000} = +66^\circ$$20'$$33.65(2)''$. We propose that this object should be referred to as PSR J0740+6620, and use this name in the rest of the article. For this pulsar we find a high $\mu_\perp$ value of $(32.6 \pm 4.1)$ mas yr$^{-1}$ that, at the NE2001 distance of $(0.68 \pm 0.10)$ kpc, makes our estimate of the intrinsic spin-down rate smaller than the observed one by almost 50\%. The corrected $\dot E_\mathrm{int}$ value above $10^{34}$ erg s$^{-1}$ and the small distance make J0740+6620 a good candidate for the detection of $\gamma$-ray pulsations, and the pulsar is actually found to coincide with the 3FGL source J0740.8+6621. The detection of $\gamma$-ray pulsations from this pulsar is presented in Section~\ref{sec:gamma}. \\

\item[$\bullet$] \textit{PSR~J0751+1807:} our estimates for the total proper motion and parallax of this pulsar differ appreciably from those reported in \citet{nss+05}, putting J0751+1807 at a greater distance of $(1.51 \pm 0.35)$ kpc. At this distance and for the transverse proper motion we measure, the Shklovskii correction to $\dot P$ is about a third of the apparent spin-down rate. The spin-down power is thus significantly smaller than that inferred from the apparent spin properties. Using the new distance and $\dot E_\mathrm{int}$ estimates, and the $\gamma$-ray energy flux for PSR~J0751+1807 reported in 2PC, we find a rather large efficiency of about 70\%, suggesting that the actual pulsar distance may be at the small end of the uncertainty range. \\

\item[$\bullet$] \textit{PSR~J0931$-$1902:} this pulsar was discovered with the Green Bank Telescope as part of the 350 MHz drift-scan survey. Its discovery and a full description of its timing properties will be presented in \citet{Lynch2016}. For this MSP, the combined Nan\c{c}ay and WSRT timing dataset allowed us to measure a modest proper motion of $\mu_\perp = (4.6 \pm 1.2)$ mas yr$^{-1}$. Even at the relatively large NE2001 distance of $(1.88 \pm 0.51)$ kpc, the observed spin-down rate $\dot P$ is found to be very weakly affected by the Shklovskii effect. Weak $\gamma$-ray pulsations are detected for this pulsar (see Section~\ref{sec:gamma}). Its high $\gamma$-ray efficiency of about 200\% (with large uncertainties) suggests a smaller distance than the NE2001-predicted value. We note that $\eta < 100$\% would imply $d < 1.4$ kpc, whereas a typical $\eta \sim 10$\% would be obtained for a much smaller distance of about 0.4 kpc. \\

\item[$\bullet$] \textit{PSR~J1231$-$1411:} at face value, the previously-published $\mu_\perp$ value of $(104 \pm 22)$ mas yr$^{-1}$ and the NE2001 distance of $(0.44 \pm 0.05)$ kpc cause the Shklovskii-corrected $\dot P_\mathrm{int}$ and $\dot E_\mathrm{int}$ terms to be negative, which is unrealistic for a rotation-powered, non-accreting MSP. Our analysis yields no parallax measurement for this pulsar, but a significantly reduced $\mu_\perp$ value of $(62.34 \pm 0.26)$ mas yr$^{-1}$ leading to $\dot E_\mathrm{int} \sim 6 \times 10^{33}$ erg s$^{-1}$, a typical value for a $\gamma$-ray MSP, and $\eta \sim 40$ \% which is also acceptable. \\

\item[$\bullet$] \textit{PSR~J1455$-$3330:} the Nan\c{c}ay timing data available for this pulsar enable us to measure the proper motion components with good accuracy, leading to $\mu_\perp = (8.11 \pm 0.05)$ mas yr$^{-1}$, lower than the previously reported $(25 \pm 12)$ mas yr$^{-1}$ from \citet{tsb+99}. Additionally, we determined a parallax of $(0.99 \pm 0.22)$ mas, placing J1455$-$3330 at a distance of $(1.01 \pm 0.22)$ kpc, and thus further away than the $(0.53 \pm 0.07)$ kpc predicted by the NE2001 model. In Section~\ref{sec:gamma} we show that J1455$-$3330 exhibits faint yet significant $\gamma$-ray pulsations. \\

\item[$\bullet$] \textit{PSR~J1614$-$2230:} our best-fit value for the transverse proper motion is consistent with earlier measurements. The detection of a significant parallax of $\pi = (1.30 \pm 0.09)$ mas enables the calculation of a revised distance to this pulsar, of $d = (0.77 \pm 0.05)$ kpc, lower than the NE2001 distance of $(1.27 \pm 0.20)$ kpc. A $\gamma$-ray efficiency of about 40\% is obtained when using the new distance and $\dot E_\mathrm{int}$ estimates, slightly higher than the $\sim 33$\% reported in 2PC, but not atypical. \\

\item[$\bullet$] \textit{PSR~J1730$-$2304:} the NE2001 model places this pulsar at a distance of $(0.53 \pm 0.05)$ kpc. We measure a parallax placing this pulsar a little farther away, at $d = (0.84 \pm 0.19)$ kpc. Using this distance value and our measurement of the transverse proper motion, we get a relatively well constrained $\dot E_\mathrm{int}$ value of $(8.4 \pm 2.2) \times 10^{32}$ erg s$^{-1}$. This $\dot E_\mathrm{int}$ value is lower than those of all known $\gamma$-ray MSPs, with the possible exceptions of PSRs~J0610$-$2100 and J1024$-$0719, for which $\dot E_\mathrm{int}$ is essentially unconstrained (see Sections \ref{sec:J0610-2100} and \ref{sec:J1024-0719}). Yet, pulsation searches in the LAT data for this MSP reveal significant $\gamma$ pulses (see Section~\ref{sec:gamma}). PSR~J1730$-$2304 may thus be the $\gamma$-ray pulsar with the lowest spin-down power value known at present. \\

\item[$\bullet$] \textit{PSR~J1741+1351:} the best-fit proper motion of $(11.62 \pm 0.13)$ mas yr$^{-1}$ is in good agreement with the value reported by \citet{Espinoza2013} of $(11.71 \pm 0.01)$ mas yr$^{-1}$. On the other hand, we do not detect the parallax signature reported in the latter article, but our 2$\sigma$ lower limit remains compatible with it. For this pulsar we therefore assume their parallax distance of $(1.08 \pm 0.05)$ kpc. For this distance and our proper motion measurement, we find a typical $\gamma$-ray efficiency of 4\% for this pulsar. \\

\item[$\bullet$] \textit{PSR~J1811$-$2405:} the detection of this pulsar in GeV $\gamma$ rays with the \textit{Fermi} LAT was reported recently by \citet{nbb+14}. Nan\c{c}ay timing measurements enable us to measure a value for the transverse proper motion of $\mu_\perp = (9.2 \pm 5.1)$ mas yr$^{-1}$, only weakly affecting the spin-down rate and spin-down power, assuming a distance for the pulsar of $(1.8 \pm 0.5)$ kpc, based on the NE2001 model. The $\gamma$-ray efficiency remains typical for this pulsar, at about 20\%, in accordance with the value published in \citet{nbb+14}. Interestingly, the 2$\sigma$ upper limit on the parallax determined from the NRT TOAs constrains the distance of PSR~J1811$-$2405 to be greater than 2.5 kpc, and thus larger than the NE2001 distance of 1.8 kpc. A greater distance would imply an increased $\gamma$-ray luminosity, and in turn, an increased efficiency. \\

% Kharchenko et al., A&A 2013 : pmra = -1.29, pmdec = -9.77
\item[$\bullet$] \textit{PSR~J1823$-$3021A:} unlike the other MSPs in our sample, this pulsar has a $\dot P$ that is a few orders of magnitude higher than those of other MSPs, and that is mostly unaffected by the Shklovskii correction. It is also the only MSP in the sample to be in a globular cluster, NGC 6624. The apparent spin-down rate is likely affected by line of sight acceleration in the cluster. Nevertheless, as was first noted by \citet{Freire2011}, the pulsar's large $\gamma$-ray luminosity indicates that it is a very energetic MSP and that most of its apparent spin-down rate is therefore intrinsic. Our proper motion measurement for PSR~J1823$-$3021A is compatible within uncertainties with the value found by \citet{Kharchenko2013} for NGC 6624, of $\sim 9.85$ kpc, suggesting that the MSP is bound to the cluster. \\

\item[$\bullet$] \textit{PSR~J2017+0603:} the transverse proper motion derived for this pulsar is small: $\mu_\perp = (2.35 \pm 0.08)$ mas yr$^{-1}$. The timing analysis also reveals a parallax signature, leading to a distance estimate for this pulsar of $(0.9 \pm 0.4)$ kpc, i.e., less than the $(1.57 \pm 0.16)$ kpc predicted by NE2001. The smaller distance and new $\dot E_\mathrm{int}$ estimate bring the $\gamma$-ray efficiency reported in 2PC from 75\% to 25\%, which is closer to the average $\eta$ value for the MSP population \citep{Johnson2013}. \\

\item[$\bullet$] \textit{PSR~J2043+1711:} for this MSP discovered at Nan\c{c}ay at the location of a \textit{Fermi} LAT source \citep{gfc+12}, the proper motion components are consistent with those reported previously and are more accurately determined: the transverse proper motion is now found to be $\mu_\perp = (12.8 \pm 0.4)$ mas yr$^{-1}$, compared to $(13 \pm 2)$ mas yr$^{-1}$) previously. We are currently insensitive to the timing parallax effect, so that our best estimate of this pulsar's distance remains the NE2001 prediction, of $(1.76 \pm 0.32)$ kpc. The 2PC luminosity leads to a $\gamma$-ray efficiency of about 100\%, suggesting a smaller distance for this pulsar. \\

\item[$\bullet$] \textit{PSR~J2214+3000:} the analysis of the Nan\c{c}ay timing data enables us to measure a parallax, placing the pulsar at $d = (0.60 \pm 0.31)$ kpc, closer than the value predicted by NE2001 based on its dispersion measure, of $(1.54 \pm 0.19)$ kpc. We also find the transverse proper motion of this pulsar to be $\mu_\perp = (20.96 \pm 0.11)$ mas yr$^{-1}$. With the new distance and revised spin-down power value, the $\gamma$-ray efficiency reported in 2PC of about 50\% decreases to a very typical $\sim 10$\%. \\

\item[$\bullet$] \textit{PSR~J2302+4442:} the timing analysis of this MSP yields a small proper motion of $\mu_\perp = (5.85 \pm 0.12)$ mas yr$^{-1}$, such that the Shklovskii correction assuming the NE2001 distance of $(1.19 \pm 0.23)$ kpc very weakly affects $\dot P$ and $\dot E$. Using the $\gamma$-ray luminosity published in 2PC of $L_\gamma \sim 6.2 \times 10^{33}$ erg s$^{-1}$, we find that the efficiency $\eta$ remains very high, at about 170\%. PSR~J2302+4442 might thus well be at a closer distance than the $(1.19 \pm 0.23)$ kpc from NE2001. 

\end{itemize}

\subsection{PSR~J0610$-$2100}
\label{sec:J0610-2100}

The $\gamma$-ray pulsations from this pulsar were first reported by \citet{Espinoza2013} and its high-energy emission properties were reassessed in 2PC. Both studies came to the conclusion that, at the NE2001 distance of $\sim 3.54$ kpc, the very low intrinsic spin-down power value is well below the empirical deathline for $\gamma$-ray emission and the inferred $\gamma$-ray efficiency is unrealistically large. Based on infrared observations in the direction of PSR~J0610$-$2100 showing pronounced nebulosity around the pulsar, \citet{Espinoza2013} proposed that unmodeled line of sight material could explain the apparently overestimated NE2001 distance. 

From the analysis of the Nan\c{c}ay timing data, we determined a transverse proper motion of $(19.10 \pm 0.08)$ mas yr$^{-1}$ that is slightly greater than the $(13 \pm 3)$ mas yr$^{-1}$ reported by \citet{bjd+06}, and the $(18.2 \pm 0.2)$ mas yr$^{-1}$ reported in \citet{Espinoza2013} and assumed in 2PC. We did not measure any significant timing parallax effect that would provide us with a revised distance estimate. Consequently, with the same distance value as used in previous $\gamma$-ray studies of this MSP and a slightly larger transverse proper motion, the issues of the very low $\dot E_\mathrm{int}$ and high $\eta$ values persist, with $\dot E_\mathrm{int} \sim 8 \times 10^{31}$ erg s$^{-1}$ and $\eta > 200$. 

\begin{figure*}[ht]
\begin{center}
\includegraphics[width= 12cm]{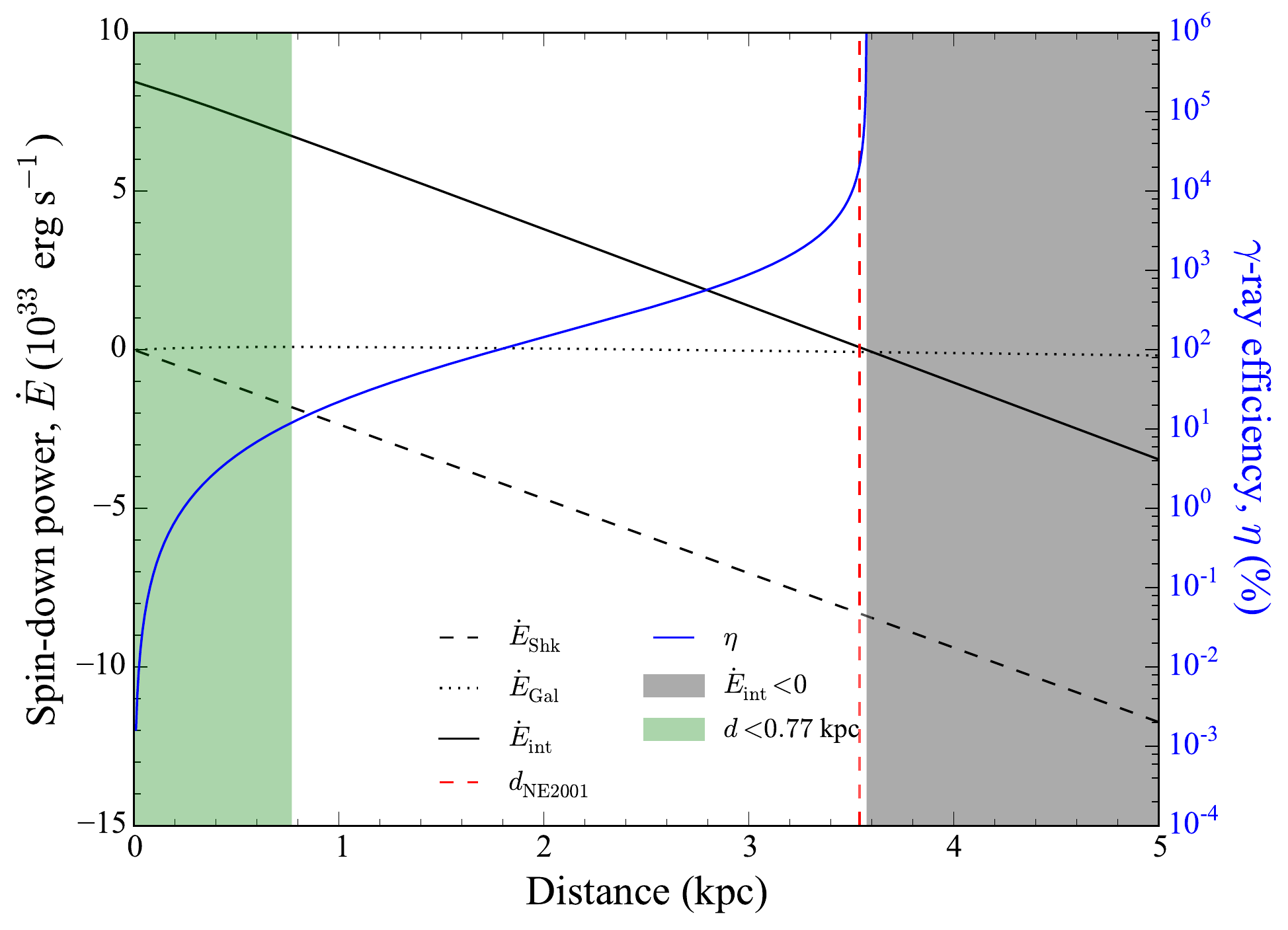}
\caption{Spin-down power $\dot E$ and $\gamma$-ray efficiency $\eta$ as a function of the distance, for PSR~J0610$-$2100. $\dot E_\mathrm{Shk}$ denotes the Shklovskii correction to $\dot E$, $\dot E_\mathrm{Gal}$ is the contribution from the line of sight acceleration in the Galactic potential, and $\dot E_\mathrm{int}$ is the spin-down power obtained after correction for these effects. The blue curve represents the $\gamma$-ray efficiency, $\eta$ (the y-axis scale is in percent), and the gray-shaded area shows the region excluded by the condition $\eta > 0$. Finally, the red dashed line shows the NE2001-predicted distance for PSR~J0610$-$2100 of 3.54 kpc, and the green-shaded region represents the zone excluded by the 2$\sigma$ lower limit on the distance from our timing analysis, of 0.77 kpc.}
\label{J0610}
\end{center}
\end{figure*}

Figure~\ref{J0610} plots the $\dot E$ and $\eta$ values for PSR~J0610$-$2100 as a function of the distance, assuming a proper motion of 19.10 mas yr$^{-1}$ and the 3FGL energy flux of $1.15 \times 10^{-11}$ erg cm$^{-2}$ s$^{-1}$. Also shown in the figure are the distance range excluded by the condition $\dot E_\mathrm{int} > 0$ (or equivalently $\eta > 0$) which is ensured for $d < 3.57$ kpc, and the range excluded by our determination of a 2$\sigma$ lower limit on the distance, of 0.77 kpc. One can immediately see that the NE2001 distance is very close to the value at which $\dot E_\mathrm{int} = 0$, hence the very low spin-down power and high efficiency. At the 2$\sigma$ lower limit on the distance the efficiency is already high, with $\eta \sim 12$\%, and it reaches 100\% at $d \sim 1.78$ kpc. 

Assuming that the observed spin-down rate is not affected by line of sight acceleration caused for instance by the presence of an unknown body attracting PSR~J0610$-$2100, we can infer that the distance to the pulsar likely lies between 0.8 and 1.8 kpc and that the pulsar's intrinsic spin-down power is in the range from $4 \times 10^{33}$ to $7 \times 10^{33}$ erg s$^{-1}$, i.e., well above the deathline for MSP $\gamma$-ray emission. In the absence of a better distance estimate for PSR~J0610$-$2100, we consider that the intrinsic spin-down power of this MSP is currently unknown and that it cannot be used for probing the $\gamma$-ray emission deathline.

\subsection{PSR~J1024$-$0719}
\label{sec:J1024-0719}

The signature of the transverse proper motion in the Nan\c{c}ay data is strong for this isolated pulsar, and the timing yields a very significant $\mu_\perp = (59.67 \pm 0.04)$ mas yr$^{-1}$, in accordance with the previously-reported measurement of \citet{vbc+09}, and among the highest values measured to date for any MSP. For this total proper motion, accounting for the Galactic acceleration and Shklovskii correction to $\dot P$ leads to a negative (hence, implausible) intrinsic spin-down rate, for any distance $d$ greater than $\sim 0.4$ kpc. As was noted by, for example, \citet{Espinoza2013} and in 2PC, using the NE2001 distance of 0.39 kpc leads to a very small intrinsic $\dot E_\mathrm{int}$ well below $10^{33}$ erg s$^{-1}$, while for the parallax distance of 0.53 kpc reported by \citet{hbo06}, it becomes negative. As can be seen from Table~\ref{tab:pm_px}, we find a parallax distance for PSR~J1024$-$0719 of $(1.13 \pm 0.18)$ kpc that is even larger than the value obtained by \citet{hbo06}, and for which the Shklovskii correction actually exceeds the apparent spin-down rate itself, thus giving a negative corrected value. In \citet{Espinoza2013} and 2PC, the issue of the negative $\gamma$-ray efficiency $\eta = L_\gamma / \dot E$ was alleviated by using the NE2001 distance. Nevertheless, our parallax measurement brings additional indication that the actual distance is likely greater than 0.4 kpc, and that another explanation than an overestimated distance needs to be found to mitigate the negative spin-down rate issue. 

As was already noted by \citet{vbc+09}, the timing of PSR~J1024$-$0719 reveals evidence for low-frequency, long-term noise in the residuals, the so-called ``red noise''. Our analysis confirms this trend: fitting for a second period derivative, $\ddot P$, yields a significant value of $(7.0 \pm 0.6) \times 10^{-32}$ s$^{-1}$. Pure magnetic dipole braking with an index $n = 2 - (P \ddot P) / \dot P^2$ such that $n = 3$ would lead to $\ddot P \sim -7 \times 10^{-38}$ s$^{-1}$, i.e., several orders of magnitude smaller than the value we measure, and with the opposite sign. Therefore, the origin of this $\ddot P$ term in the timing residuals of PSR~J1024$-$0719 is likely extrinsic. 

A natural explanation for the negative Shklovskii-corrected spin-down rate and the presence of a significant second period derivative in the Nan\c{c}ay timing data could be that PSR~J1024$-$0719 undergoes acceleration along the line of sight, caused by the existence of a second body in the vicinity of the pulsar. \citet{Joshi1997} presented a method for determining the mass of the putative companion and the orbital parameters, in the cases where only a small fraction of the orbit is covered. With this method, the full determination of the orbital parameters and the mass requires the measurement of the first five derivatives of the period. With the present data for J1024$-$0719, only $P$, $\dot P$, and $\ddot P$ can be significantly detected. Subsequent period derivatives may become measurable with an increased timing dataset. 

Nonetheless, deep VLT observations of the field of PSR~J1024$-$0719 conducted by \citet{Sutaria2003} revealed the presence of two stars near the position of the pulsar. One possibility could therefore be that PSR~J1024$-$0719 is associated with one of the two nearby stars, in a wide orbit causing the MSP to undergo acceleration along our line of sight. Follow-up observations by \citet{Bassa2016} may confirm the association of the pulsar with either the bright or the faint optical source, or deny this scenario. 

We conclude that the intrinsic spin-down rate ($\dot P_\mathrm{int}$) of this MSP is not accurately known at present, and that consequently the pulsar should not be used as a probe of the deathline for $\gamma$-ray emission from MSPs.

\section{Gamma-ray analysis}
\label{sec:gamma}

We characterized the $\gamma$-ray emission from PSRs J0740+6620, J0931$-$1902, J1455$-$3330, and J1730$-$2304 by analyzing data from the Large Area Telescope (LAT), the electron-positron pair conversion telescope on the \textit{Fermi} satellite launched in June 2008 \citep{Atwood2009}. We selected LAT data from the much improved Pass 8 reconstruction algorithms \citep{Atwood2013}. The event list was restricted to those recorded between 2008 August 4 and 2015 July 1, with energies between 0.1 and 300 GeV, and with zenith angles smaller than 90$^\circ$ to limit the contamination of the datasets from the Earth's limb. The analysis was carried out with the \textit{Fermi} Science Tools\footnote{See http://fermi.gsfc.nasa.gov/ssc/data/analysis/scitools/overview.html} (STs) v10-01-01. The data were phase-folded using the ephemerides obtained from the analysis described in Section~\ref{sec:radiotiming}, with the \textit{Fermi} plug-in for \textsc{Tempo2} \citep{Ray2011}. 

For each of the four MSPs we created individual $\gamma$-ray datasets by selecting photons found within 15$^\circ$ of the pulsars. In parallel, we constructed spectral models for these regions of interest (ROIs) by selecting 3FGL sources within 20$^\circ$ of the MSPs, and by including models representing the Galactic diffuse emission and the isotropic diffuse and residual instrumental background emission, using the gll\_iem\_v06.fits and iso\_P8R2\_SOURCE\_V6\_v06.txt files produced by the \textit{Fermi} LAT collaboration. Of the four MSPs considered in this high-energy analysis, PSRs J0740+6620 and J0931$-$1902 have counterparts in the 3FGL catalog, named 3FGL~J0740.8+6621 and J0930.9$-$1904, respectively. We shifted the positions of these two sources in our models to the radio timing positions found for the MSPs. For PSRs J1455$-$3330 and J1730$-$2304 we added new sources, also placed at the best-fit coordinates from the timing analysis. 

The spectral parameters of sources more than 5$^\circ$ away from the pulsars were fixed at the values determined in the 3FGL analysis, except for sources with Test Statistic (TS) values higher than 1000, which were fixed at the 3FGL results if more than 10$^\circ$ away. Spectral parameters of other sources were left free in the fits. The contributions from the four MSPs were modeled as a function of energy $E$ as exponentially cutoff power laws of the form $N_0 \left( E / \mathrm{1\ GeV} \right)^{-\Gamma} \exp \left( -E / E_c \right)$, where $N_0$ represents a normalization factor, $\Gamma$ is the power law index and $E_c$ is the exponential cutoff energy. The \textit{Fermi} ST \textsc{gtlike} was used in conjunction with the MINUIT optimizer to determine the spectral parameters of the sources in the models, by means of a maximum likelihood analysis. To increase the signal-to-noise ratios of the pulsars, we restricted the datasets to pulse phase ranges determined from the inspection of the $\gamma$-ray pulse profiles, described below.

The results of the spectral analysis for PSRs J0740+6620, J0931$-$1902, J1455$-$3330, and J1730$-$2304 are presented in Table~\ref{tab:gamma}. All four MSPs are detected as significant sources of $\gamma$-ray emission, as can be noted from the TS values. The $\gamma$-ray emission of two of the four pulsars is very weak, which prevented us from measuring the $\Gamma$ parameter in these cases. To derive meaningful constraints on the other parameters, the spectral index of PSR~J0931$-$1902 was fixed at the value of $1.85 \pm 0.16$ found in 3FGL. For PSR~J1455$-$3330 we fixed the $\Gamma$ parameter to 1.3, the average value of MSP spectral indices tabulated in 2PC. Also given in Table~\ref{tab:gamma} are the $\gamma$-ray luminosities above 0.1 GeV, $L_\gamma = 4 \pi h d^2$ (this assumes a beaming correction factor, $f_\Omega$, defined in, e.g., 2PC, of 1), and the conversion efficiencies $\eta = L_\gamma / \dot E_\mathrm{int}$. The efficiency values for PSRs J0740+6620 and J1455$-$3330 are fairly typical for $\gamma$-ray MSPs. For PSRs J0931$-$1902 and J1730$-$2304 the large efficiency uncertainties stem mainly from the distance uncertainties, and the efficiencies are consistent with being below 100\%. We also tried fitting the pulsar spectra with simple power laws of the form $N_0 \left( E / \mathrm{1\ GeV} \right)^{-\Gamma}$, and found that the exponentially cutoff power-law shapes are preferred in all four cases, with 1 to 4$\sigma$ significance. 

\begin{table*}
\caption{
Properties of PSRs J0740+6620, J0931$-$1902, J1455$-$3330, and J1730$-$2304 in GeV $\gamma$ rays. The spectral indices of PSRs J0931$-$1902 and J1455$-$3330 were fixed at the values listed in the Table. These values are marked with a star. Details on the parameters and on the analysis method can be found in Section~\ref{sec:gamma}. Quoted error bars are the $1\sigma$ statistical uncertainties. 
}
\label{tab:gamma}
\centering
\begin{tabular}{l c c c c}
\hline
\hline
Parameter & J0740+6620 & J0931$-$1902 & J1455$-$3330 & J1730$-$2304 \\
\hline
Selected phase range & $[0.08; 0.7]$ & $[0;0.25] \cup [0.9;1]$ & $[0.2; 0.5]$ & $[0.3; 0.5]$ \\
Source TS & 159.6 & 66.3 & 48.8 & 91.5 \\
Spectral index, $\Gamma$ & $1.3 \pm 0.5$ & $1.85^\star$ & $1.3^\star$ & $2.5 \pm 0.2$ \\
Cutoff energy, $E_c$ (GeV) & $2.6 \pm 1.5$ & $6.5 \pm 3.4$ & $1.9 \pm 0.5$ & $7 \pm 5$ \\
Photon flux above 0.1 GeV ($10^{-8}$ cm$^{-2}$ s$^{-1}$) & $0.3 \pm 0.1$ & $0.16 \pm 0.03$ & $0.17 \pm 0.04$ & $2.4 \pm 0.6$ \\
Energy flux above 0.1 GeV, $h$ ($10^{-12}$ erg cm$^{-2}$ s$^{-1}$) & $3.6 \pm 2.3$ & $6.7 \pm 2.2$ & $4.2 \pm 1.5$ & $10 \pm 5$ \\
Luminosity, $L_\gamma = 4 \pi h d^2$ ($10^{32}$ erg s$^{-1}$) & $2.0 \pm 1.4$ & $28 \pm 18$ & $5 \pm 3$ & $8 \pm 5$ \\
Efficiency, $\eta = L_\gamma / \dot E_\mathrm{int}$ (\%) & $1.7 \pm 1.3$ & $200 \pm 130$ & $28 \pm 16$ & $100 \pm 80$ \\
Weighted H-test TS (significance) & 162.3 (11.2$\sigma$) & 50.1 (6.0$\sigma$) & 77.6 (7.6$\sigma$) & 36.8 (5.1$\sigma$) \\
\hline
\end{tabular}
\end{table*}

\begin{figure*}[ht]
\begin{center}
\includegraphics[width=8cm]{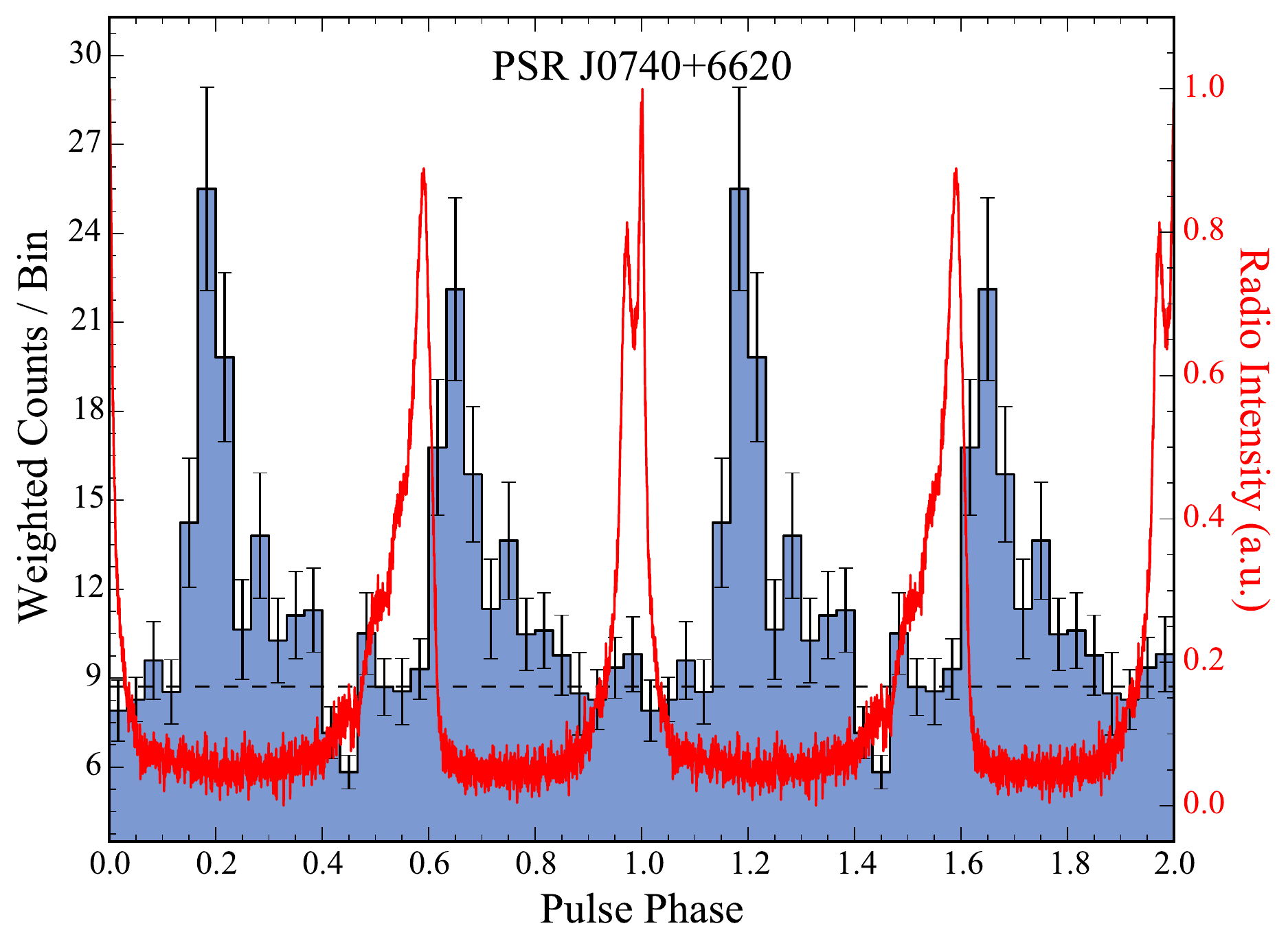}
\includegraphics[width=8cm]{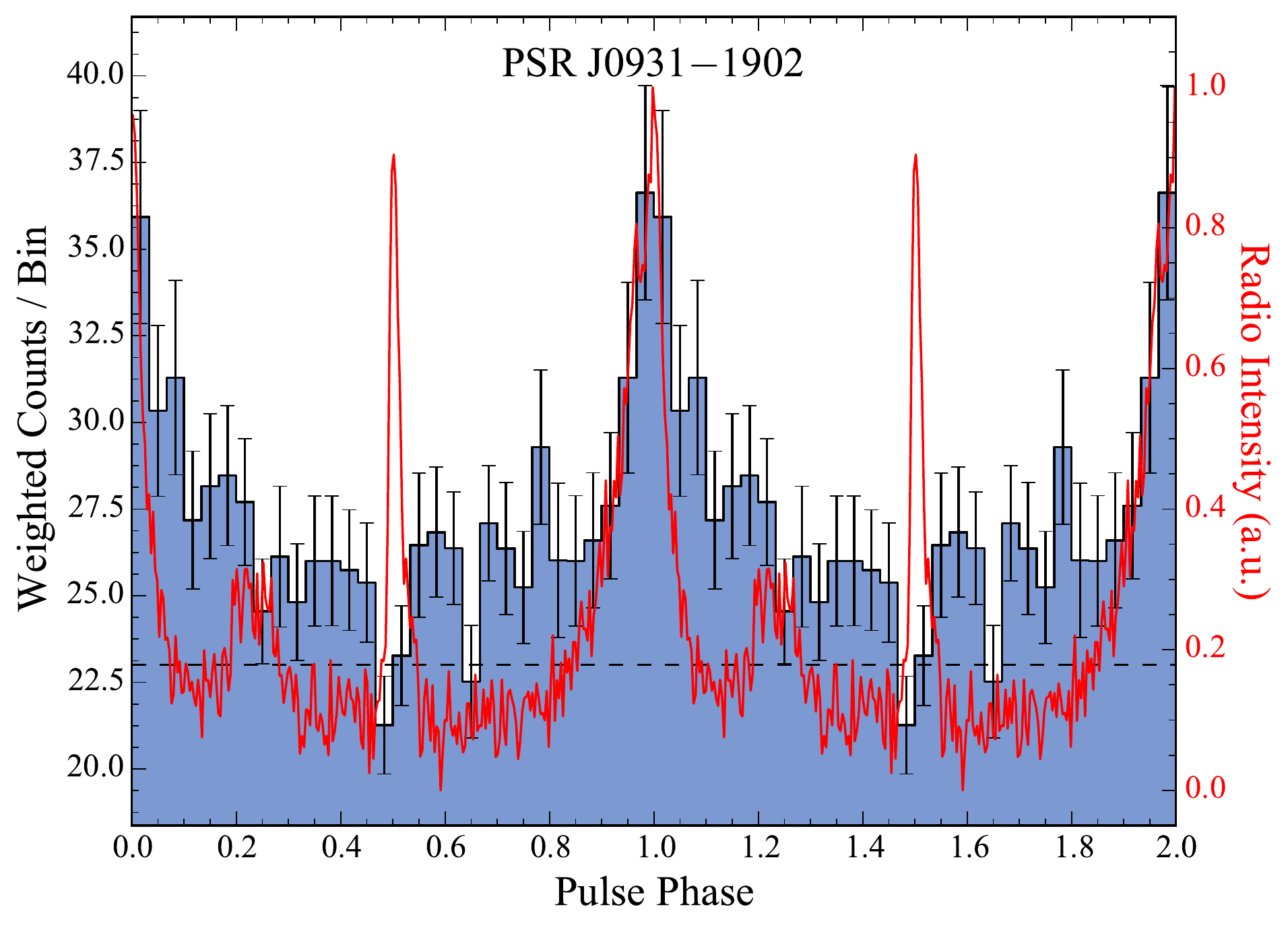}
\includegraphics[width=8cm]{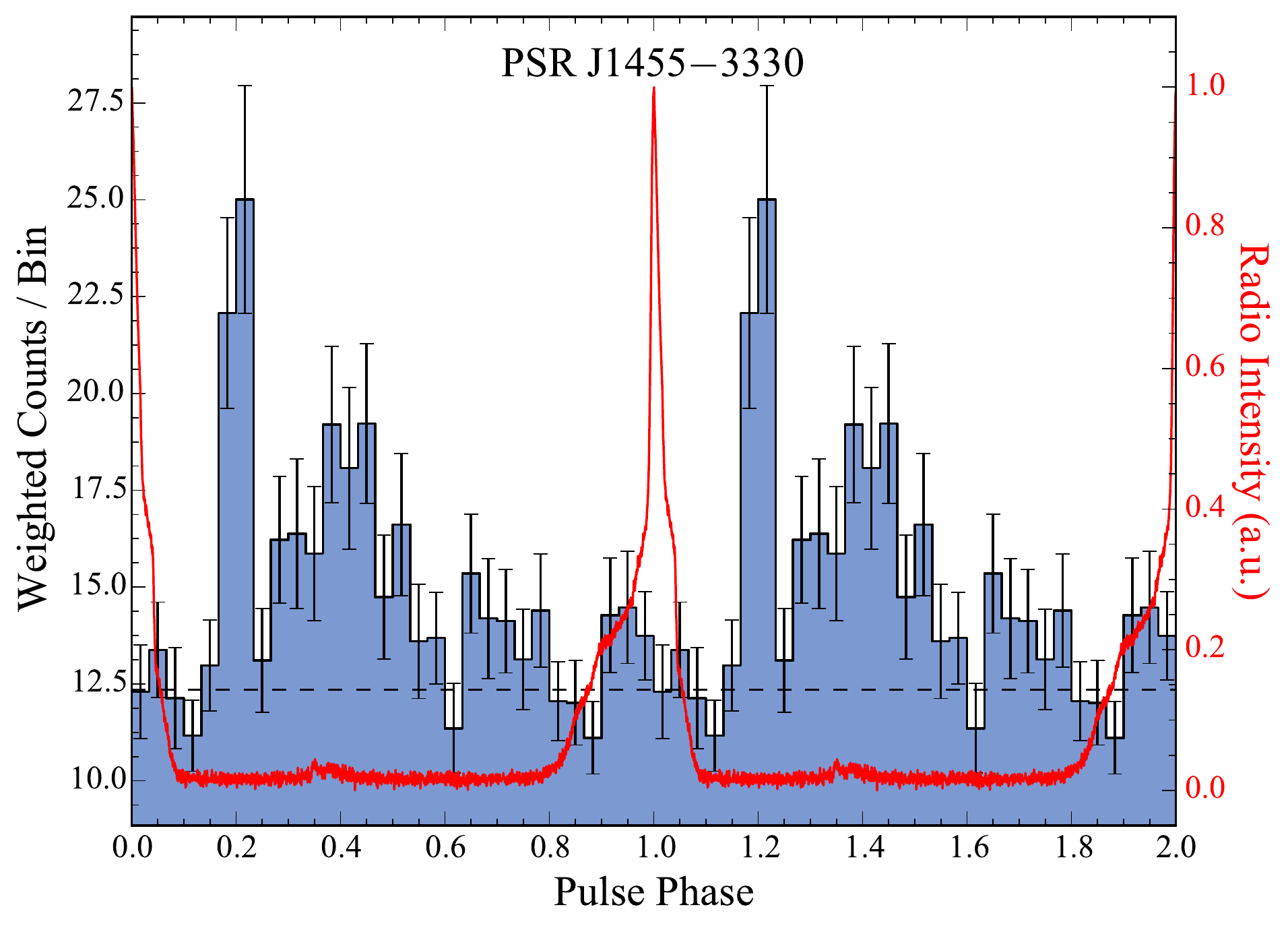}
\includegraphics[width=8cm]{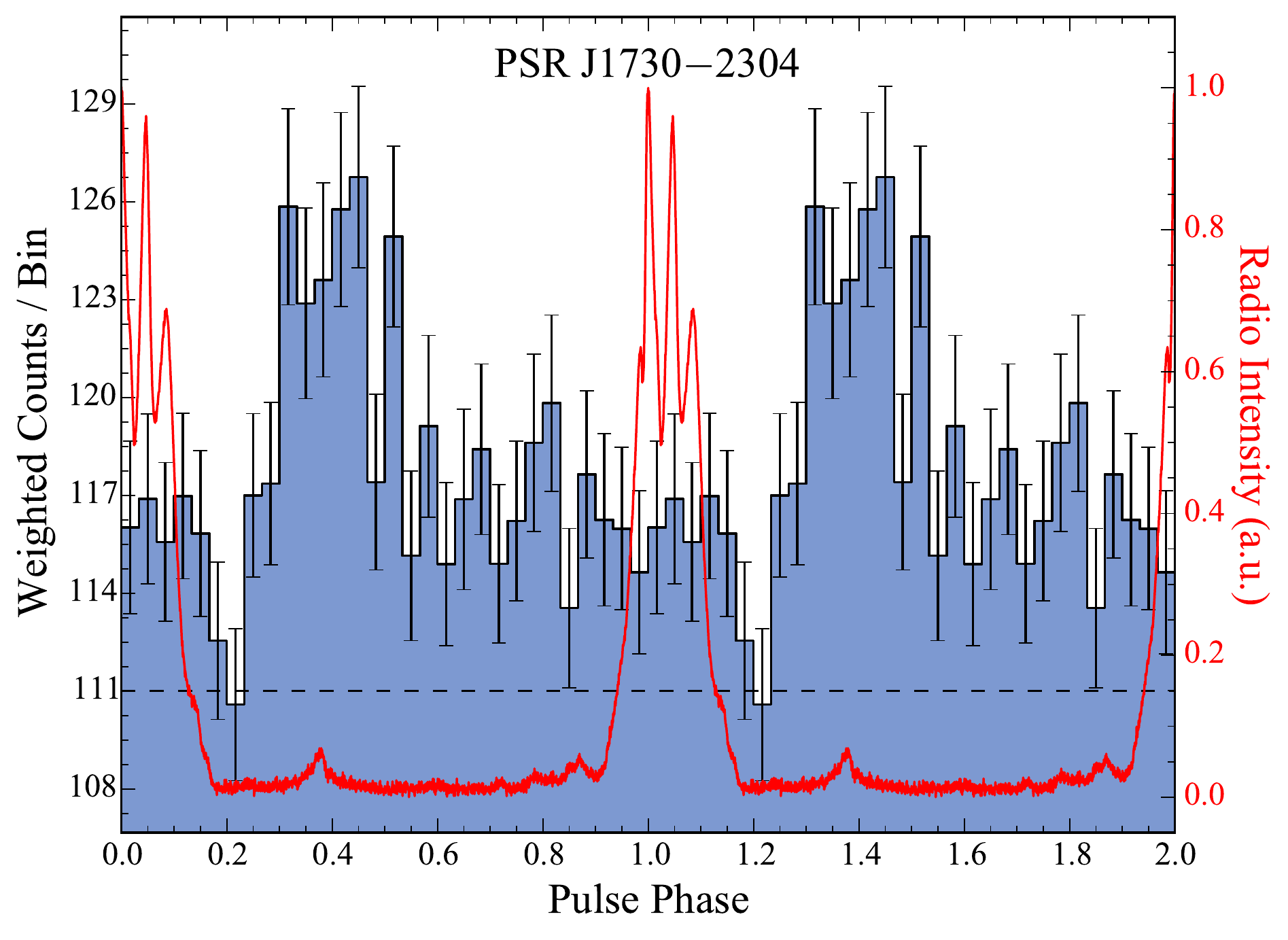}
\caption{Integrated \textit{Fermi} LAT $\gamma$-ray and Nan\c{c}ay radio profiles for the MSPs J0740+6620, J0931$-$1902, J1455$-$3330 and J1730$-$2304. Two full radio profiles are shown. For PSR~J0740+6620 we display the LAT events recorded after MJD 55800. Radio light curves all correspond to 1.4~GHz Nan\c{c}ay profiles recorded with the NUPPI backend, with 2048 bins per rotation except for PSR~J0931$-$1902 for which the number of bins was reduced to 256 to improve the signal-to-noise ratio of the profile. The 30-bin $\gamma$-ray light curves were constructed by selecting \textit{Fermi} LAT photons found within 5$^\circ$ of the pulsars and with energies above 0.1 GeV. The photons were then weighted by the probability that they were emitted by the pulsars, as calculated based on spectral likelihood results.}
\label{phasos}
\end{center}
\end{figure*}

The best spectral models for the regions around the MSPs were used to compute probabilities that the photons in the ROIs were emitted by the pulsars, using the ST \textsc{gtsrcprob}. Having assigned weights and pulse phases to each photon in our datasets, we produced weighted $\gamma$-ray profiles; these profiles are shown in Figure~\ref{phasos}. We find weighted H-test TS significances \citep{Kerr2011} above $5\sigma$ for all four objects, indicating significant $\gamma$-ray detections and bringing the number of MSPs detected as pulsed $\gamma$-ray sources to 71. For PSR~J0740+6620 our timing solution is found to determine accurate pulse phases after MJD 55800; we therefore selected events recorded during this time interval. Figure~\ref{phasos} also shows integrated 1.4~GHz radio profiles, with the correct relative radio/$\gamma$-ray alignment. In all cases the first $\gamma$-ray peak lags the main radio component, a fairly standard feature in radio and $\gamma$-ray pulsars.

\section{Discussion}

\begin{figure*}[ht]
\begin{center}
\includegraphics[width=9cm]{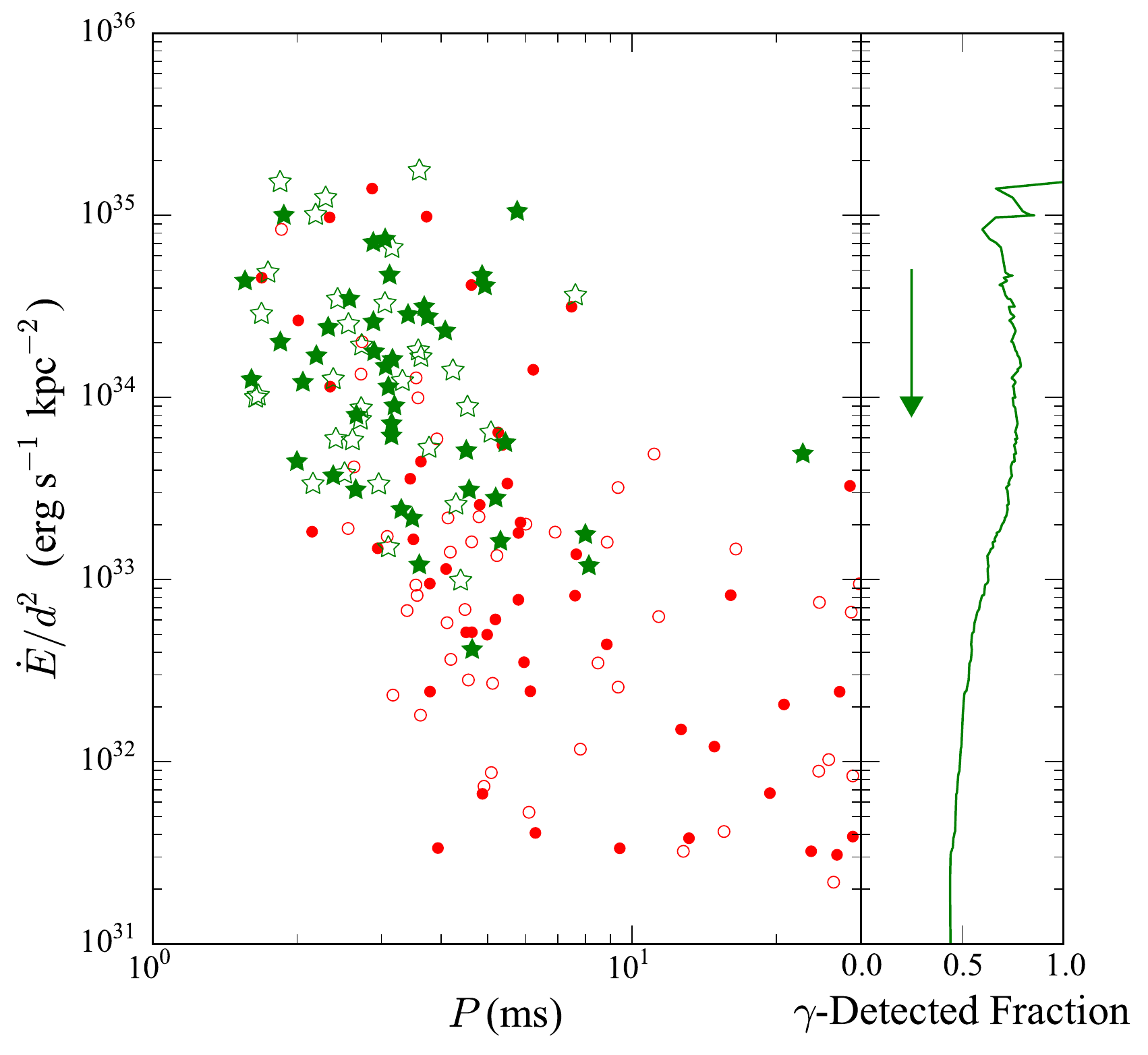}
\includegraphics[width=9cm]{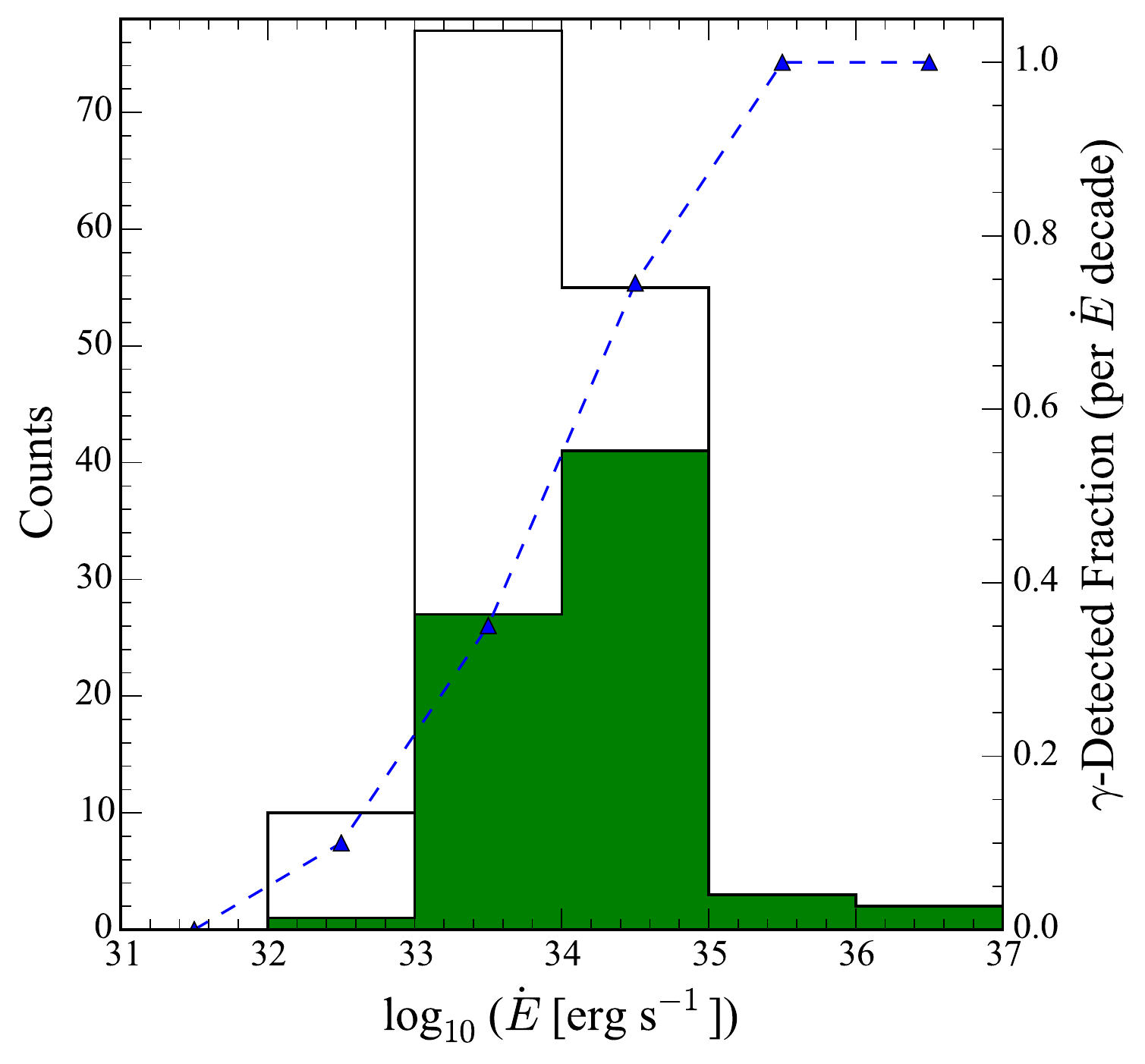}
\caption{\textit{Left:} Spin-down power $\dot E$ divided by the square of the distance $d$ as a function of the period $P$, for MSPs in the Galactic disk. PSRs J1823$-$3021A and J1824$-$2452A, two MSPs in globular clusters but detected in $\gamma$ rays, are included in the plot. Green stars represent $\gamma$-ray MSPs, undetected ones are shown as red circles. \citet{GuillemotTauris2014} explain non-detections of energetic and distant MSPs in $\gamma$ rays as due to unfavorable viewing angles. All $\dot E$ values are corrected for the effect of the acceleration in the Galactic potential. Filled symbols indicate pulsars for which we could correct for the kinematic Shklovskii effect. The right-hand panel shows the cumulative fraction of MSPs detected in $\gamma$ rays, with decreasing $\dot E / d^2$ as indicated by the green arrow. \textit{Right:} Spin-down power values for the MSPs with $\dot E / d^2 \geq 1.5 \times 10^{32}$ erg s$^{-1}$ kpc$^{-2}$. Half of the MSPs in this sample are seen with the \textit{Fermi} LAT. The green histogram shows the $\gamma$-detected MSPs, the empty histogram corresponds to the total number of MSPs in each $\dot E$ decade. The dashed line shows the fraction of $\gamma$-detected MSPs per $\dot E$ decade.}
\label{edotd2}
\end{center}
\end{figure*}

With the \textit{Fermi} LAT detections of PSRs J0740+6620, J0931$-$1902, J1455$-$3330 and J1730$-$2304, 71 MSPs have now been observed to emit GeV $\gamma$-ray pulsations. Nearly half of the total number of known $\gamma$-ray pulsars are MSPs, and about a third of all Galactic disk MSPs (i.e., MSPs outside of globular clusters) are seen in $\gamma$ rays. Figure~\ref{edotd2}a is an update of the spin-down power $\dot E$ normalized by $d^2$ versus $P$ plot for Galactic disk MSPs with measured spin-down rates, previously shown in \citet{GuillemotTauris2014}, but with a larger MSP sample and with the new $\gamma$-ray detections. The two known $\gamma$-ray MSPs in globular clusters PSRs J1823$-$3021A and J1824$-$2452A are included in the plot. On the other hand, the $\gamma$-ray pulsars J0610$-$2100 and J1024$-$0719 are not plotted, having implausible Shklovskii-corrected spin-down powers (see Table~\ref{tab:MSPs}), as argued in Sections~\ref{sec:J0610-2100} and \ref{sec:J1024-0719}.

One striking feature of the MSP population, as can be seen from Figure~\ref{edotd2}a, is that a large majority of the energetic and nearby ones are seen in $\gamma$ rays. Above $\dot E / d^2 = 5 \times 10^{33}$ erg s$^{-1}$ kpc$^{-2}$, 75\% of Galactic disk MSPs with known spin-down rates have been detected by the \textit{Fermi} LAT. High $\dot E / d^2$ MSPs that are undetected in $\gamma$ rays could be further away than currently estimated, they could be less energetic (in particular if their proper motions and thus their Shklovskii-corrected spin-down rates are unknown), or they could be seen under unfavorable viewing angles as argued by \citet{GuillemotTauris2014}. At present, the least energetic $\gamma$-ray MSP known is PSR~J1730$-$2304, with $\dot E_\mathrm{int} = (8.4 \pm 2.2) \times 10^{32}$ erg s$^{-1}$. The latter value thus represents the current empirical deathline for $\gamma$-ray emission from MSPs, and future LAT observations of MSPs will tell if lower-$\dot E$ MSPs can produce detectable $\gamma$-ray emission. 

Figure~\ref{edotd2}b shows a histogram of $\dot E_\mathrm{int}$ values for MSPs with $\dot E_\mathrm{int} / d^2 \geq 1.5 \times 10^{32}$ erg s$^{-1}$ kpc$^{-2}$, being the limit above which 50\% of the MSPs shown in Figure~\ref{edotd2}a are seen with the LAT. To a first approximation, the MSPs in this sample can be considered \textit{detectable}. One possible improvement would consist of accounting for the background $\gamma$-ray emission present at the locations of these MSPs, and comparing these background levels to the maximum $\gamma$-ray fluxes one could expect from the MSPs. Nevertheless, we find that the majority of these MSPs are located at high Galactic latitudes and therefore generally lie in regions of weak background $\gamma$-ray emission. The plot confirms the idea that the $\gamma$-ray detectability of an MSP depends crucially on its spin-down power. In this sample, the ratio of $\gamma$-detected MSPs increases with $\dot E$, and in particular 100\% of the MSPs with $\dot E \geq 10^{35}$ erg s$^{-1}$ are seen by the LAT.

\begin{figure*}[ht]
\begin{center}
\includegraphics[width=15cm]{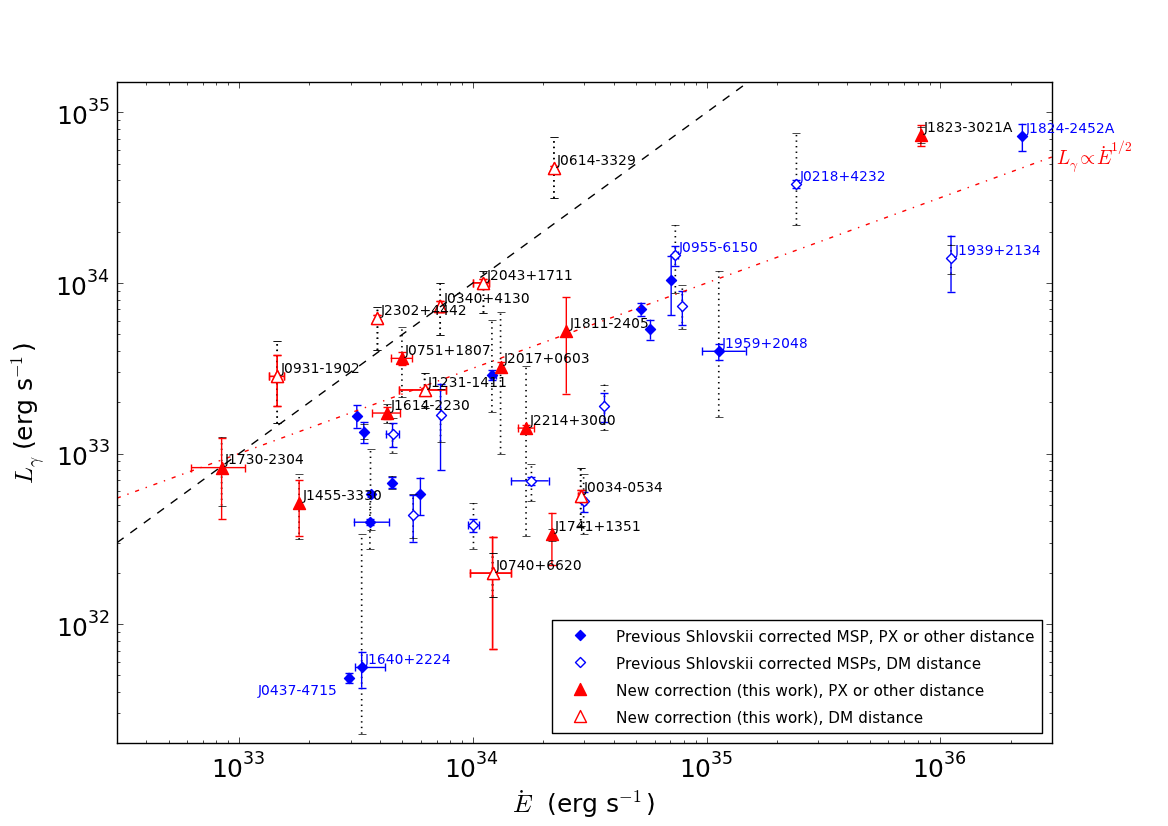}
\caption{Luminosity $L_\gamma = 4 \pi h d^2$ above 0.1 GeV as a function of the spin-down power ($\dot E$) for the sample of MSPs considered in this work (red triangles), and other MSPs with Shklovskii-corrected $\dot E$ values (blue diamonds). Vertical error bars in gray represent the uncertainties due to the distance, while colored error bars represent the uncertainties on the $\gamma$-ray energy flux, $h$. The dashed line represents $L_\gamma = \dot E$, and the dash-dotted line indicates the heuristic luminosity $L_\gamma^h = \sqrt{10^{33} \dot E}$. Empty symbols represent pulsars with distance values estimated via the dispersion measure and the NE2001 model of \citet{NE2001}, filled symbols are pulsars whose distances were determined with other methods, such as the measurement of the timing parallax.}
\label{lgammaedot}
\end{center}
\end{figure*}

Figure~\ref{lgammaedot} confirms that MSPs are detectable in $\gamma$ rays only when $\dot E > \dot E_\mathrm{death} \sim 10^{33}$ erg s$^{-1}$. Our discovery of $\gamma$-ray pulsations from PSR~J1730$-$2304 suggests that the minimum $\dot E_\mathrm{death}$ value could be less than that. Models describing particle acceleration and $\gamma$-ray emission in the magnetosphere of pulsars must therefore be able to explain high-energy emission from such low $\dot E$ pulsars. In addition, with this reduced empirical deathline we expect the Milky Way to host more $\gamma$-ray-emitting MSPs than previously thought. Population synthesis analyses aiming to predict the contribution from unresolved MSPs to the diffuse $\gamma$-ray emission in the Milky Way and their possible contribution to the Galactic Center excess \citep[see for example][]{Calore2014} will need to account for this increased number of $\gamma$-ray MSPs in the Galaxy. We note that all of the MSPs with $L_\gamma \gtrsim \dot E$ in Figure~\ref{lgammaedot} have distances estimated via their DM values and the NE2001 model, while MSPs with distances determined with other methods generally have efficiencies well below 100\%. Given the large DM distance uncertainties discussed in Section~\ref{sec:pulsarsec}, the $\gamma$-ray luminosities greater than $\dot E$ in Figure~\ref{lgammaedot} do not necessarily indicate that $\dot E$ is underestimated. The moment of inertia $I = \dfrac{2}{5} M R^2 = 10^{45}$ g cm$^2$ (where $M$ and $R$ are the neutron star mass and radius) that we use is compatible with the range of most observed neutron star masses, $1 < M < 2$ M$_\odot$, combined with the currently acceptable range of predicted neutron star radii, $9 < R < 15$ km. 

One key ingredient in several of these population studies is the relationship between the $\gamma$-ray luminosity of MSPs and the spin-down power. Knowing how $L_\gamma$ scales with $\dot E$, one can in principle predict the emitted flux and, after populating the Galaxy with MSPs, estimate their contribution to the diffuse emission. Precise proper motion and distance measurements such as those described in Section~\ref{sec:radioresults} help refine this relationship. Quite striking in Figure~\ref{lgammaedot} is that for $\dot E \gtrsim \dot E_\mathrm{death}$, the luminosity is mostly uncorrelated with $\dot E$. For instance, for $\dot E$ values between $10^{33}$ and $10^{34}$ erg s$^{-1}$, calculated $L_\gamma$ values are found to vary by two orders of magnitude. The apparent lack of a clear correlation contrasts with what is seen for young pulsars with higher spin-down power, which follow the rough $\sqrt{\dot E}$ trend suggested in Figure~\ref{lgammaedot} for $\dot E \gtrsim 5 \times 10^{34}$ erg s$^{-1}$. The $\sqrt{\dot E}$ relation comes from simple arguments \citep{Arons1996} that only partially describe the accelerating region.

The luminosity values shown in Figure~\ref{lgammaedot} were calculated as $L_\gamma = 4 \pi h d^2$, i.e., assuming a geometrical correction factor, $f_\Omega$, of 1. The spread in the luminosity distribution could thus be partly explained by our line of sight sampling the $\gamma$-ray beam's neutron star latitude profile more or less favorably.  However, \citet{Johnson2014} found little variation in the $f_\Omega$ factors obtained for a sample of $\gamma$-ray MSPs and under different emission models, their $f_\Omega$ values being typically close to unity. Varying geometrical correction factors thus likely play a limited role in the large $L_\gamma$ spread.  The spin-down power may also be affected by magnetospheric parameters, such as the magnetic inclination, $\alpha$, or the current flows. \citet{Spitkovsky2006} and \citet{Petri2012} considered pulsars with force-free magnetospheres and found the following expression for the spin-down power: $\dot E_\mathrm{ff} \simeq 3/2 \dot E_\mathrm{vac} \left(1 + \sin^2 \alpha\right)$ where $\dot E_\mathrm{vac} = 4 \pi^2 I \dot P / P^3$ is the vacuum spin-down power typically used for estimating $\dot E$. Similar to the correction to $L_\gamma$ due to the $f_\Omega$ term, this correction to $\dot E$ can only partially mitigate the spread. 

How brightly an MSP emits in $\gamma$ rays, and how much of its total energy budget it converts into high-energy emission, surely depends on the shape and extent of the zone where electron cascades occur, and on the electric potential that can be sustained across the zone. The latter is mitigated by the plasma currents flowing through and around the zone. Continued modeling efforts to reproduce observations such as in Figure~\ref{lgammaedot}, and especially to allow predictions of the $\gamma$-ray luminosity for arbitrary $P$, $\dot P$, and $\alpha$ values would permit improved estimates of the MSP contribution to the diffuse background.

\section{Summary}
\label{sec:summary}

We have presented the analysis of several years of Nan\c{c}ay and Westerbork radio timing data for a selection of $\gamma$-ray MSPs, which enabled us to determine their proper motions, and measure timing parallaxes for four of them. These parameters were used to improve our estimates of their spin-down power values by correcting for the Shlovskii effect, and of their $\gamma$-ray luminosities. We have also presented the analysis of more than six years of Pass 8 \textit{Fermi} LAT $\gamma$-ray data, leading to the discovery of high-energy pulsations for four MSPs: PSRs J0740+6620, J0931$-$1902, J1455$-$3330, and J1730$-$2304. The latter object is now the least energetic $\gamma$-ray pulsar known, setting the empirical deathline for $\gamma$-ray emission from MSPs to $\dot E_\mathrm{death} \sim 8 \times 10^{32}$ erg s$^{-1}$. PSRs J0610$-$2100 and J1024$-$0719, whose $\dot E$ values are likely unknown, could be even less energetic objects. 

By considering the population of known Galactic disk MSPs, we have confirmed that those seen to emit $\gamma$ rays by the \textit{Fermi} LAT are the energetic and nearby ones. In the sample of MSPs with $\dot E / d^2$ values above $5 \times 10^{33}$ erg s$^{-1}$ kpc$^{-2}$, 75\% are observed to emit pulsed $\gamma$-ray emission. Nevertheless, selecting $\gamma$-ray MSPs with Shklovskii-corrected $\dot E$ values, we have shown that above $\dot E_\mathrm{death}$ the spin-down power and the $\gamma$-ray luminosity appear mostly uncorrelated, in spite of the improved $\dot E$ and $L_\gamma$ estimates. Varying moments of inertia, emission geometries and more realistic prescriptions for the energy budget that MSPs can convert into $\gamma$-ray emission could mitigate the lack of apparent correlation. Continued analyses of Pass 8 LAT data may also reveal $\gamma$-ray pulsations from even less energetic MSPs, constraining the $\gamma$-ray emission deathline and the spin-down-power versus luminosity relationship further.

\begin{acknowledgements}

We thank Cees Bassa for helpful discussions and constructive suggestions. 

The \textit{Fermi} LAT Collaboration acknowledges generous ongoing support
from a number of agencies and institutes that have supported both the
development and the operation of the LAT as well as scientific data analysis.
These include the National Aeronautics and Space Administration and the
Department of Energy in the United States, the Commissariat \`a l'Energie Atomique
and the Centre National de la Recherche Scientifique / Institut National de Physique
Nucl\'eaire et de Physique des Particules in France, the Agenzia Spaziale Italiana
and the Istituto Nazionale di Fisica Nucleare in Italy, the Ministry of Education,
Culture, Sports, Science and Technology (MEXT), High Energy Accelerator Research
Organization (KEK) and Japan Aerospace Exploration Agency (JAXA) in Japan, and
the K.~A.~Wallenberg Foundation, the Swedish Research Council and the
Swedish National Space Board in Sweden.

Additional support for science analysis during the operations phase is gratefully 
acknowledged from the Istituto Nazionale di Astrofisica in Italy and the Centre 
National d'\'Etudes Spatiales in France.

The Nan\c{c}ay Radio Observatory is operated by the Paris Observatory, associated 
with the French Centre National de la Recherche Scientifique (CNRS).

The Westerbork Synthesis Radio Telescope is operated by the Netherlands Institute 
for Radio Astronomy (ASTRON) with support from The Netherlands Foundation for 
Scientific Research (NWO).

\end{acknowledgements}

\bibliographystyle{aa}
\bibliography{NancayFermiPaper}

\begin{thebibliography}{52}
\expandafter\ifx\csname natexlab\endcsname\relax\def\natexlab#1{#1}\fi

\bibitem[{{Abdo} {et~al.}(2013){Abdo}, {Ajello}, {Allafort}, {Baldini},
  {Ballet}, {Barbiellini}, {Baring}, {Bastieri}, {Belfiore}, {Bellazzini}, \&
  et~al.}]{Fermi2PC}
{Abdo}, A.~A., {Ajello}, M., {Allafort}, A., {et~al.} 2013, \apjs, 208, 17

\bibitem[{{Acero} {et~al.}(2015){Acero}, {Ackermann}, {Ajello}, {Albert},
  {Atwood}, {Axelsson}, {Baldini}, {Ballet}, {Barbiellini}, {Bastieri},
  {Belfiore}, {Bellazzini}, {Bissaldi}, {Blandford}, {Bloom}, {Bogart},
  {Bonino}, {Bottacini}, {Bregeon}, {Britto}, {Bruel}, {Buehler}, {Burnett},
  {Buson}, {Caliandro}, {Cameron}, {Caputo}, {Caragiulo}, {Caraveo},
  {Casandjian}, {Cavazzuti}, {Charles}, {Chaves}, {Chekhtman}, {Cheung},
  {Chiang}, {Chiaro}, {Ciprini}, {Claus}, {Cohen-Tanugi}, {Cominsky}, {Conrad},
  {Cutini}, {D'Ammando}, {de Angelis}, {DeKlotz}, {de Palma}, {Desiante},
  {Digel}, {Di Venere}, {Drell}, {Dubois}, {Dumora}, {Favuzzi}, {Fegan},
  {Ferrara}, {Finke}, {Franckowiak}, {Fukazawa}, {Funk}, {Fusco}, {Gargano},
  {Gasparrini}, {Giebels}, {Giglietto}, {Giommi}, {Giordano}, {Giroletti},
  {Glanzman}, {Godfrey}, {Grenier}, {Grondin}, {Grove}, {Guillemot}, {Guiriec},
  {Hadasch}, {Harding}, {Hays}, {Hewitt}, {Hill}, {Horan}, {Iafrate}, {Jogler},
  {J{\'o}hannesson}, {Johnson}, {Johnson}, {Johnson}, {Johnson}, {Kamae},
  {Kataoka}, {Katsuta}, {Kuss}, {La Mura}, {Landriu}, {Larsson}, {Latronico},
  {Lemoine-Goumard}, {Li}, {Li}, {Longo}, {Loparco}, {Lott}, {Lovellette},
  {Lubrano}, {Madejski}, {Massaro}, {Mayer}, {Mazziotta}, {McEnery},
  {Michelson}, {Mirabal}, {Mizuno}, {Moiseev}, {Mongelli}, {Monzani},
  {Morselli}, {Moskalenko}, {Murgia}, {Nuss}, {Ohno}, {Ohsugi}, {Omodei},
  {Orienti}, {Orlando}, {Ormes}, {Paneque}, {Panetta}, {Perkins},
  {Pesce-Rollins}, {Piron}, {Pivato}, {Porter}, {Racusin}, {Rando}, {Razzano},
  {Razzaque}, {Reimer}, {Reimer}, {Reposeur}, {Rochester}, {Romani},
  {Salvetti}, {S{\'a}nchez-Conde}, {Saz Parkinson}, {Schulz}, {Siskind},
  {Smith}, {Spada}, {Spandre}, {Spinelli}, {Stephens}, {Strong}, {Suson},
  {Takahashi}, {Takahashi}, {Tanaka}, {Thayer}, {Thayer}, {Thompson},
  {Tibaldo}, {Tibolla}, {Torres}, {Torresi}, {Tosti}, {Troja}, {Van Klaveren},
  {Vianello}, {Winer}, {Wood}, {Wood}, \& {Zimmer}}]{Fermi3FGL}
{Acero}, F., {Ackermann}, M., {Ajello}, M., {et~al.} 2015, \apjs, 218, 23

\bibitem[{{Alpar} {et~al.}(1982){Alpar}, {Cheng}, {Ruderman}, \&
  {Shaham}}]{Alpar1982}
{Alpar}, M.~A., {Cheng}, A.~F., {Ruderman}, M.~A., \& {Shaham}, J. 1982, \nat,
  300, 728

\bibitem[{{Arons}(1996)}]{Arons1996}
{Arons}, J. 1996, \aaps, 120, C49

\bibitem[{{Atwood} {et~al.}(2013){Atwood}, {Albert}, {Baldini}, {Tinivella},
  {Bregeon}, {Pesce-Rollins}, {Sgr{\`o}}, {Bruel}, {Charles}, {Drlica-Wagner},
  {Franckowiak}, {Jogler}, {Rochester}, {Usher}, {Wood}, {Cohen-Tanugi}, \&
  {S.~Zimmer for the Fermi-LAT Collaboration}}]{Atwood2013}
{Atwood}, W., {Albert}, A., {Baldini}, L., {et~al.} 2013,
  arXiv:astro-ph/1303.3514

\bibitem[{{Atwood} {et~al.}(2009){Atwood}, {Abdo}, {Ackermann}, {Althouse},
  {Anderson}, {Axelsson}, {Baldini}, {Ballet}, {Band}, {Barbiellini}, \&
  et~al.}]{Atwood2009}
{Atwood}, W.~B., {Abdo}, A.~A., {Ackermann}, M., {et~al.} 2009, \apj, 697, 1071

\bibitem[{{Bassa} {et~al.}(2016)}]{Bassa2016}
{Bassa}, C.~G. {et~al.} 2016, in preparation

\bibitem[{{Bhalerao} \& {Kulkarni}(2011)}]{bk11}
{Bhalerao}, V.~B. \& {Kulkarni}, S.~R. 2011, \apjl, 737, L1

\bibitem[{{Bhattacharya} \& {van den Heuvel}(1991)}]{Bhattacharya1991}
{Bhattacharya}, D. \& {van den Heuvel}, E.~P.~J. 1991, \physrep, 203, 1

\bibitem[{{Burgay} {et~al.}(2006){Burgay}, {Joshi}, {D'Amico}, {Possenti},
  {Lyne}, {Manchester}, {McLaughlin}, {Kramer}, {Camilo}, \& {Freire}}]{bjd+06}
{Burgay}, M., {Joshi}, B.~C., {D'Amico}, N., {et~al.} 2006, \mnras, 368, 283

\bibitem[{{Calore} {et~al.}(2014){Calore}, {Di Mauro}, \&
  {Donato}}]{Calore2014}
{Calore}, F., {Di Mauro}, M., \& {Donato}, F. 2014, \apj, 796, 14

\bibitem[{{Cordes} \& {Lazio}(2002)}]{NE2001}
{Cordes}, J.~M. \& {Lazio}, T.~J.~W. 2002, arXiv:astro-ph/0207156

\bibitem[{{Demorest} {et~al.}(2010){Demorest}, {Pennucci}, {Ransom}, {Roberts},
  \& {Hessels}}]{dpr+10}
{Demorest}, P.~B., {Pennucci}, T., {Ransom}, S.~M., {Roberts}, M.~S.~E., \&
  {Hessels}, J.~W.~T. 2010, \nat, 467, 1081

\bibitem[{{Desvignes} {et~al.}(2016)}]{Desvignes2016}
{Desvignes}, G. {et~al.} 2016, submitted to MNRAS

\bibitem[{{Espinoza} {et~al.}(2013){Espinoza}, {Guillemot}, {{\c C}elik},
  {Weltevrede}, {Stappers}, {Smith}, {Kerr}, {Zavlin}, {Cognard}, {Eatough},
  {Freire}, {Janssen}, {Camilo}, {Desvignes}, {Hewitt}, {Hou}, {Johnston},
  {Keith}, {Kramer}, {Lyne}, {Manchester}, {Ransom}, {Ray}, {Shannon},
  {Theureau}, \& {Webb}}]{Espinoza2013}
{Espinoza}, C.~M., {Guillemot}, L., {{\c C}elik}, {\"O}., {et~al.} 2013,
  \mnras, 430, 571

\bibitem[{{Folkner} {et~al.}(2009){Folkner}, {Williams}, \&
  {Boggs}}]{Folkner2009}
{Folkner}, W.~M., {Williams}, J.~G., \& {Boggs}, D.~H. 2009, IPN Progress
  Report, 42-178

\bibitem[{{Freire} {et~al.}(2011){Freire}, {Abdo}, {Ajello}, {Allafort},
  {Ballet}, {Barbiellini}, {Bastieri}, {Bechtol}, {Bellazzini}, {Blandford},
  {Bloom}, {Bonamente}, {Borgland}, {Brigida}, {Bruel}, {Buehler}, {Buson},
  {Caliandro}, {Cameron}, {Camilo}, {Caraveo}, {Cecchi}, {{\c C}elik},
  {Charles}, {Chekhtman}, {Cheung}, {Chiang}, {Ciprini}, {Claus}, {Cognard},
  {Cohen-Tanugi}, {Cominsky}, {de Palma}, {Dermer}, {do Couto e Silva},
  {Dormody}, {Drell}, {Dubois}, {Dumora}, {Espinoza}, {Favuzzi}, {Fegan},
  {Ferrara}, {Focke}, {Fortin}, {Fukazawa}, {Fusco}, {Gargano}, {Gasparrini},
  {Gehrels}, {Germani}, {Giglietto}, {Giordano}, {Giroletti}, {Glanzman},
  {Godfrey}, {Grenier}, {Grondin}, {Grove}, {Guillemot}, {Guiriec}, {Hadasch},
  {Harding}, {J{\'o}hannesson}, {Johnson}, {Johnson}, {Johnston}, {Katagiri},
  {Kataoka}, {Keith}, {Kerr}, {Kn{\"o}dlseder}, {Kramer}, {Kuss}, {Lande},
  {Latronico}, {Lee}, {Lemoine-Goumard}, {Longo}, {Loparco}, {Lovellette},
  {Lubrano}, {Lyne}, {Manchester}, {Marelli}, {Mazziotta}, {McEnery},
  {Michelson}, {Mizuno}, {Moiseev}, {Monte}, {Monzani}, {Morselli},
  {Moskalenko}, {Murgia}, {Nakamori}, {Nolan}, {Norris}, {Nuss}, {Ohsugi},
  {Okumura}, {Omodei}, {Orlando}, {Ozaki}, {Paneque}, {Parent},
  {Pesce-Rollins}, {Pierbattista}, {Piron}, {Porter}, {Rain{\`o}}, {Ransom},
  {Ray}, {Reimer}, {Reimer}, {Reposeur}, {Ritz}, {Romani}, {Roth},
  {Sadrozinski}, {Saz Parkinson}, {Shannon}, {Siskind}, {Smith}, {Spinelli},
  {Stappers}, {Suson}, {Takahashi}, {Tanaka}, {Tauris}, {Thayer}, {Theureau},
  {Thompson}, {Thorsett}, {Tibaldo}, {Torres}, {Tosti}, {Troja},
  {Vandenbroucke}, {Van Etten}, {Vasileiou}, {Venter}, {Vianello}, {Vilchez},
  {Vitale}, {Waite}, {Wang}, {Wood}, {Yang}, {Ziegler}, \&
  {Zimmer}}]{Freire2011}
{Freire}, P.~C.~C., {Abdo}, A.~A., {Ajello}, M., {et~al.} 2011, Science, 334,
  1107

\bibitem[{{Guillemot} {et~al.}(2012){Guillemot}, {Freire}, {Cognard},
  {Johnson}, {Takahashi}, {Kataoka}, {Desvignes}, {Camilo}, {Ferrara},
  {Harding}, {Janssen}, {Keith}, {Kerr}, {Kramer}, {Parent}, {Ransom}, {Ray},
  {Saz Parkinson}, {Smith}, {Stappers}, \& {Theureau}}]{gfc+12}
{Guillemot}, L., {Freire}, P.~C.~C., {Cognard}, I., {et~al.} 2012, \mnras, 422,
  1294

\bibitem[{{Guillemot} \& {Tauris}(2014)}]{GuillemotTauris2014}
{Guillemot}, L. \& {Tauris}, T.~M. 2014, \mnras, 439, 2033

\bibitem[{{Hobbs} {et~al.}(2005){Hobbs}, {Lorimer}, {Lyne}, \&
  {Kramer}}]{hllk05}
{Hobbs}, G., {Lorimer}, D.~R., {Lyne}, A.~G., \& {Kramer}, M. 2005, \mnras,
  360, 974

\bibitem[{{Hobbs} {et~al.}(2006){Hobbs}, {Edwards}, \& {Manchester}}]{TEMPO2}
{Hobbs}, G.~B., {Edwards}, R.~T., \& {Manchester}, R.~N. 2006, \mnras, 369, 655

\bibitem[{{Hotan} {et~al.}(2006){Hotan}, {Bailes}, \& {Ord}}]{hbo06}
{Hotan}, A.~W., {Bailes}, M., \& {Ord}, S.~M. 2006, \mnras, 369, 1502

\bibitem[{{Hotan} {et~al.}(2004){Hotan}, {van Straten}, \&
  {Manchester}}]{PSRCHIVE}
{Hotan}, A.~W., {van Straten}, W., \& {Manchester}, R.~N. 2004, \pasa, 21, 302

\bibitem[{{Hou} {et~al.}(2014){Hou}, {Smith}, {Guillemot}, {Cheung}, {Cognard},
  {Craig}, {Espinoza}, {Johnston}, {Kramer}, {Reimer}, {Reposeur}, {Shannon},
  {Stappers}, \& {Weltevrede}}]{Hou2014}
{Hou}, X., {Smith}, D.~A., {Guillemot}, L., {et~al.} 2014, \aap, 570, A44

\bibitem[{{Huber}(1981)}]{Huber1981}
{Huber}, P.~J. 1981, {Robust statistics}

\bibitem[{{Johnson} {et~al.}(2013){Johnson}, {Guillemot}, {Kerr}, {Cognard},
  {Ray}, {Wolff}, {B{\'e}gin}, {Janssen}, {Romani}, {Venter}, {Grove},
  {Freire}, {Wood}, {Cheung}, {Casandjian}, {Stairs}, {Camilo}, {Espinoza},
  {Ferrara}, {Harding}, {Johnston}, {Kramer}, {Lyne}, {Michelson}, {Ransom},
  {Shannon}, {Smith}, {Stappers}, {Theureau}, \& {Thorsett}}]{Johnson2013}
{Johnson}, T.~J., {Guillemot}, L., {Kerr}, M., {et~al.} 2013, \apj, 778, 106

\bibitem[{{Johnson} {et~al.}(2014){Johnson}, {Venter}, {Harding}, {Guillemot},
  {Smith}, {Kramer}, {{\c C}elik}, {den Hartog}, {Ferrara}, {Hou}, {Lande}, \&
  {Ray}}]{Johnson2014}
{Johnson}, T.~J., {Venter}, C., {Harding}, A.~K., {et~al.} 2014, \apjs, 213, 6

\bibitem[{{Joshi} \& {Rasio}(1997)}]{Joshi1997}
{Joshi}, K.~J. \& {Rasio}, F.~A. 1997, \apj, 479, 948

\bibitem[{{Karuppusamy} {et~al.}(2008){Karuppusamy}, {Stappers}, \& {van
  Straten}}]{Karuppusamy2008}
{Karuppusamy}, R., {Stappers}, B., \& {van Straten}, W. 2008, \pasp, 120, 191

\bibitem[{{Kerr}(2011)}]{Kerr2011}
{Kerr}, M. 2011, ApJ, 732, 38

\bibitem[{{Kharchenko} {et~al.}(2013){Kharchenko}, {Piskunov}, {Schilbach},
  {R{\"o}ser}, \& {Scholz}}]{Kharchenko2013}
{Kharchenko}, N.~V., {Piskunov}, A.~E., {Schilbach}, E., {R{\"o}ser}, S., \&
  {Scholz}, R.-D. 2013, \aap, 558, A53

\bibitem[{{Kuulkers} {et~al.}(2003){Kuulkers}, {den Hartog}, {in't Zand},
  {Verbunt}, {Harris}, \& {Cocchi}}]{Kuulkers2003}
{Kuulkers}, E., {den Hartog}, P.~R., {in't Zand}, J.~J.~M., {et~al.} 2003,
  \aap, 399, 663

\bibitem[{{Lorimer} \& {Kramer}(2012)}]{handbook}
{Lorimer}, D.~R. \& {Kramer}, M. 2012, {Handbook of Pulsar Astronomy}

\bibitem[{{Lutz} \& {Kelker}(1973)}]{LutzKelker1973}
{Lutz}, T.~E. \& {Kelker}, D.~H. 1973, \pasp, 85, 573

\bibitem[{{Lynch} {et~al.}(2016)}]{Lynch2016}
{Lynch}, R.~S. {et~al.} 2016, in preparation

\bibitem[{{Manchester} {et~al.}(2005){Manchester}, {Hobbs}, {Teoh}, \&
  {Hobbs}}]{ATNF}
{Manchester}, R.~N., {Hobbs}, G.~B., {Teoh}, A., \& {Hobbs}, M. 2005, \aj, 129,
  1993

\bibitem[{{Matthews} {et~al.}(2015){Matthews}, {Nice}, {Fonseca},
  {Arzoumanian}, {Crowter}, {Demorest}, {Dolch}, {Ellis}, {Ferdman},
  {Gonzalez}, {Jones}, {Jones}, {Lam}, {Levin}, {McLaughlin}, {Pennucci},
  {Ransom}, {Stairs}, {Stovall}, {Swiggum}, \& {Zhu}}]{Matthews2015}
{Matthews}, A.~M., {Nice}, D.~J., {Fonseca}, E., {et~al.} 2015,
  arXiv:astro-ph/1509.08982

\bibitem[{{Ng} {et~al.}(2014){Ng}, {Bailes}, {Bates}, {Bhat}, {Burgay},
  {Burke-Spolaor}, {Champion}, {Coster}, {Johnston}, {Keith}, {Kramer},
  {Levin}, {Petroff}, {Possenti}, {Stappers}, {van Straten}, {Thornton},
  {Tiburzi}, {Bassa}, {Freire}, {Guillemot}, {Lyne}, {Tauris}, {Shannon}, \&
  {Wex}}]{nbb+14}
{Ng}, C., {Bailes}, M., {Bates}, S.~D., {et~al.} 2014, \mnras, 439, 1865

\bibitem[{{Nice} {et~al.}(2005){Nice}, {Splaver}, {Stairs}, {L{\"o}hmer},
  {Jessner}, {Kramer}, \& {Cordes}}]{nss+05}
{Nice}, D.~J., {Splaver}, E.~M., {Stairs}, I.~H., {et~al.} 2005, \apj, 634,
  1242

\bibitem[{{P{\'e}tri}(2012)}]{Petri2012}
{P{\'e}tri}, J. 2012, \mnras, 424, 605

\bibitem[{{Ransom} {et~al.}(2011){Ransom}, {Ray}, {Camilo}, {Roberts}, {{\c
  C}elik}, {Wolff}, {Cheung}, {Kerr}, {Pennucci}, {DeCesar}, {Cognard}, {Lyne},
  {Stappers}, {Freire}, {Grove}, {Abdo}, {Desvignes}, {Donato}, {Ferrara},
  {Gehrels}, {Guillemot}, {Gwon}, {Harding}, {Johnston}, {Keith}, {Kramer},
  {Michelson}, {Parent}, {Saz Parkinson}, {Romani}, {Smith}, {Theureau},
  {Thompson}, {Weltevrede}, {Wood}, \& {Ziegler}}]{rrc+11}
{Ransom}, S.~M., {Ray}, P.~S., {Camilo}, F., {et~al.} 2011, \apjl, 727, L16

\bibitem[{{Ray} {et~al.}(2011){Ray}, {Kerr}, {Parent}, {Abdo}, {Guillemot},
  {Ransom}, {Rea}, {Wolff}, {Makeev}, {Roberts}, {Camilo}, {Dormody}, {Freire},
  {Grove}, {Gwon}, {Harding}, {Johnston}, {Keith}, {Kramer}, {Michelson},
  {Romani}, {Saz Parkinson}, {Thompson}, {Weltevrede}, {Wood}, \&
  {Ziegler}}]{Ray2011}
{Ray}, P.~S., {Kerr}, M., {Parent}, D., {et~al.} 2011, \apjs, 194, 17

\bibitem[{{Reardon} {et~al.}(2016){Reardon}, {Hobbs}, {Coles}, {Levin},
  {Keith}, {Bailes}, {Bhat}, {Burke-Spolaor}, {Dai}, {Kerr}, {Lasky},
  {Manchester}, {Os{\l}owski}, {Ravi}, {Shannon}, {van Straten}, {Toomey},
  {Wang}, {Wen}, {You}, \& {Zhu}}]{Reardon2016}
{Reardon}, D.~J., {Hobbs}, G., {Coles}, W., {et~al.} 2016, \mnras, 455, 1751

\bibitem[{{Shklovskii}(1970)}]{Shklovskii1970}
{Shklovskii}, I.~S. 1970, \sovast, 13, 562

\bibitem[{{Spitkovsky}(2006)}]{Spitkovsky2006}
{Spitkovsky}, A. 2006, \apjl, 648, L51

\bibitem[{{Stovall} {et~al.}(2014){Stovall}, {Lynch}, {Ransom}, {Archibald},
  {Banaszak}, {Biwer}, {Boyles}, {Dartez}, {Day}, {Ford}, {Flanigan}, {Garcia},
  {Hessels}, {Hinojosa}, {Jenet}, {Kaplan}, {Karako-Argaman}, {Kaspi},
  {Kondratiev}, {Leake}, {Lorimer}, {Lunsford}, {Martinez}, {Mata},
  {McLaughlin}, {Roberts}, {Rohr}, {Siemens}, {Stairs}, {van Leeuwen},
  {Walker}, \& {Wells}}]{Stovall2014}
{Stovall}, K., {Lynch}, R.~S., {Ransom}, S.~M., {et~al.} 2014, \apj, 791, 67

\bibitem[{{Sutaria} {et~al.}(2003){Sutaria}, {Ray}, {Reisenegger}, {Hertling},
  {Quintana}, \& {Minniti}}]{Sutaria2003}
{Sutaria}, F.~K., {Ray}, A., {Reisenegger}, A., {et~al.} 2003, \aap, 406, 245

\bibitem[{{Tauris} \& {van den Heuvel}(2006)}]{Tauris2006}
{Tauris}, T.~M. \& {van den Heuvel}, E.~P.~J. 2006, {Formation and evolution of
  compact stellar X-ray sources}, ed. W.~H.~G. {Lewin} \& M.~{van der Klis},
  623--665

\bibitem[{{Taylor}(1992)}]{Taylor1992}
{Taylor}, J.~H. 1992, Philosophical Transactions of the Royal Society of London
  Series A, 341, 117

\bibitem[{{Toscano} {et~al.}(1999){Toscano}, {Sandhu}, {Bailes}, {Manchester},
  {Britton}, {Kulkarni}, {Anderson}, \& {Stappers}}]{tsb+99}
{Toscano}, M., {Sandhu}, J.~S., {Bailes}, M., {et~al.} 1999, \mnras, 307, 925

\bibitem[{{Verbiest} {et~al.}(2009){Verbiest}, {Bailes}, {Coles}, {Hobbs}, {van
  Straten}, {Champion}, {Jenet}, {Manchester}, {Bhat}, {Sarkissian}, {Yardley},
  {Burke-Spolaor}, {Hotan}, \& {You}}]{vbc+09}
{Verbiest}, J.~P.~W., {Bailes}, M., {Coles}, W.~A., {et~al.} 2009, \mnras, 400,
  951

\bibitem[{{Verbiest} {et~al.}(2012){Verbiest}, {Weisberg}, {Chael}, {Lee}, \&
  {Lorimer}}]{Verbiest2012}
{Verbiest}, J.~P.~W., {Weisberg}, J.~M., {Chael}, A.~A., {Lee}, K.~J., \&
  {Lorimer}, D.~R. 2012, \apj, 755, 39

\end{thebibliography}

\end{document}